# General Semiparametric Shared Frailty Model: Estimation and Simulation with frailtySurv

**John V. Monaco**
Naval Postgraduate School

**Malka Gorfine**
Tel Aviv University

**Li Hsu**
Fred Hutchinson
Cancer Research Center

## Abstract

The R package **frailtySurv** for simulating and fitting semi-parametric shared frailty models is introduced. Package **frailtySurv** implements semi-parametric consistent estimators for a variety of frailty distributions, including gamma, log-normal, inverse Gaussian and power variance function, and provides consistent estimators of the standard errors of the parameters' estimators. The parameters' estimators are asymptotically normally distributed, and therefore statistical inference based on the results of this package, such as hypothesis testing and confidence intervals, can be performed using the normal distribution. Extensive simulations demonstrate the flexibility and correct implementation of the estimator. Two case studies performed with publicly available datasets demonstrate applicability of the package. In the Diabetic Retinopathy Study, the onset of blindness is clustered by patient, and in a large hard drive failure dataset, failure times are thought to be clustered by the hard drive manufacturer and model.

## 1. Introduction

The semi-parametric Cox proportional hazards (PH) regression model was developed by Sir David Cox (1972) and is by far the most popular model for survival analysis. The model defines a *hazard function*, which is the rate of an event occurring at any given time, given the observation is still at risk, as a function of the observed covariates. When data consist of independent and identically distributed observations, the parameters of the Cox PH model are estimated using the partial likelihood (Cox 1975) and the Breslow (1974) estimator.

Often, the assumption of independent and identically distributed observations is violated. In clinical data, it is typical for survival times to be clustered or depend on some unobserved co-variates. This can be due to geographical clustering, subjects sharing common genes, or some other predisposition that cannot be observed directly. Survival times can also be clustered by subject when there are multiple observations per subject with common baseline hazard. For example, the Diabetic Retinopathy Study was conducted to determine the time to the onset of blindness in high risk diabetic patients and to evaluate the effectiveness of laser treatment. The treatment was administered to one randomly-selected eye in each patient, leaving the other eye untreated. Obviously, the two eyes' measurements of each patient are clustered by patient due to unmeasured patient-specific effects.

Clustered survival times are not limited to clinical data. Computer components often exhibit clustering due to different materials and manufacturing processes. The failure rate of magnetic storage devices is of particular interest since component failure can result in data loss. A large backup storage provider may utilize tens of thousands of hard drives consisting of hundreds of different hard drive models. In evaluating the time until a hard drive becomes inoperable, it is important to consider operating conditions as well as the hard drive model. Hard drive survival times depend on the model since commercial grade models may be built out of better materials and designed to have longer lifetimes than consumer grade models. The above two examples are used in Section 5 for demonstrating the usage of the **frailtySurv** package (Monaco, Gorfine, and Hsu 2018).

Clayton (1978) accounted for cluster-specific unobserved effects by introducing a random effect term into the proportional hazards model, which later became known as the *shared frailty model*. A shared frailty model includes a latent random variable, the *frailty*, which comprises the unobservable dependency between members of a cluster. The frailty has a multiplicative effect on the hazard, and given the observed covariates and unobserved frailty, the survival times within a cluster are assumed independent.

Under the shared frailty model, the hazard function at time $t$ of observation $j$ of cluster $i$ is given by

$$\lambda_{ij}\left(t|\mathbf{Z}_{ij},\omega_i\right) = \omega_i \lambda_0\left(t\right) e^{\beta^\top \mathbf{Z}_{ij}}, \quad j=1,\ldots,m_i, \quad i=1,\ldots,n, \tag{1}$$

where $\omega_i$ is an unobservable frailty variate of cluster $i$, $\lambda_0(t)$ is the unknown common baseline hazard function, $\beta$ is the unknown regression coefficient vector, and $\mathbf{Z}_{ij}$ is the observed vector of covariates of observation $j$ in cluster $i$. The frailty variates $\omega_1,\ldots,\omega_n$, are independent and identically distributed with known density $f\left(\cdot;\theta\right)$ and unknown parameter $\theta$.

There are currently several estimation techniques available with a corresponding R package (R Core Team 2018) for fitting a shared frailty model, as shown in Table 1. In a parametric model, the baseline hazard function is of known parametric form, with several unknown parameters. Parameter estimation of parametric models is performed by the maximum marginal likelihood (MML) approach (Duchateau and Janssen 2007; Wienke 2010). The **parfm** package (Munda, Rotolo, and Legrand 2012) implements several parametric frailty models. In a semi-parametric model, the baseline hazard function is left unspecified, a highly important feature, as often in practice the shape of the baseline hazard function is unknown. Under the semi-parametric setting, the top downloaded packages, **survival** (Therneau 2018b) and **coxme** (Therneau 2018a), implement the penalized partial likelihood (PPL). **frailtypack** parameter estimates are obtained by nonlinear least squares (NLS) with the hazard function and cumulative hazard function modeled by a 4th order cubic M-spline and integrated M-spline, respectively (Rondeau, Mazroui, and Gonzalez 2012). Since the frailty term is a latent variable, expectation maximization (EM) is also a natural estimation strategy for semi-parametric models, implemented by **phmm** (Donohue and Xu 2017). More recently, a hierarchical-likelihood (h-likelihood, or HL) method (Do Ha, Lee, and kee Song 2001) has been used to fit hierarchical shared frailty models, implemented by **frailtyHL** (Do Ha, Noh, and Lee 2018). Both R packages **R2BayesX** (Brezger, Kneib, and Lang 2005; Gu 2014) and **gss** (Hirsch and Wienke 2012) can fit a shared frailty model and support only Gaussian random effects with the baseline hazard function estimated by penalized splines.

| `package::function` | $\lambda_0$ | Estimation procedure | Frailty distributions | Weekly downloads |
|---|---|---|---|---|
| **survival**`::coxph` | NP | PPL | Gamma, LN, LT | 3905 |
| **gss**`::sscox` | NP | PPL | LN | 1120 |
| **coxme**`::coxme` | NP | PPL | LN | 260 |
| **frailtypack**`::frailtyPenal` | NP | NLS | Gamma, LN | 98 |
| **R2BayesX**`::bayesx` | NP | PPL | LN | 58 |
| **phmm**`::phmm` | NP | EM | LN | 52 |
| **frailtySurv**`::fitfrail` | NP | PFL | Gamma, LN, IG, PVF | 50 |
| **frailtyHL**`::frailtyHL` | NP | HL | Gamma, LN | 50 |
| **parfm**`::parfm` | P | MML | Gamma, PS, IG | 49 |
| **survBayes**`::survBayes` | NP | Bayes | Gamma, LN | 28 |

Table 1: R functions for fitting shared frailty models. NP = nonparametric, P = parametric, PPL = penalized partial likelihood, NLS = nonlinear least squares, EM = expectation maximization, PFL = pseudo full likelihood, HL = h-likelihood, MML = maximum marginal likelihood, LN = log-normal, LT = log-t, IG = inverse Gaussian, PS = positive stable. Weekly downloads are averages from the time the package first appears on the RStudio CRAN mirror through 2016-06-01, as reported by the RStudio CRAN package download logs: http://cran-logs.rstudio.com/.

This work introduces the **frailtySurv** R package, an implementation of Gorfine, Zucker, and Hsu (2006) and Zucker, Gorfine, and Hsu (2008), wherein an estimation procedure for semi-parametric shared frailty models with general frailty distributions was proposed. Gorfine *et al.* (2006) addresses some limitations of other existing methods. Specifically, all other available semi-parametric packages can only be applied with gamma, log-normal (LN), and log-t (LT) frailty distributions. In contrast, the semi-parametric estimation procedure used in **frailtySurv** supports general frailty distributions with finite moments, and the current version of **frailtySurv** implements gamma, log-normal, inverse Gaussian (IG), and power variance function (PVF) frailty distributions. Additionally, the asymptotic properties of most of the semi-parametric estimators in Table 1 are not known. In contrast, the regression coefficients' estimators, the frailty distribution parameter estimator, and the baseline hazard estimator of **frailtySurv** are backed by a rigorous large-sample theory (Gorfine *et al.* 2006; Zucker *et al.* 2008). In particular, these estimators are consistent and asymptotically normally distributed. A consistent covariance-matrix estimator of the regression coefficients' estimators and the frailty distribution parameter's estimator is provided by Gorfine *et al.* (2006) and Zucker *et al.* (2008), also implemented by **frailtySurv**. Alternatively, **frailtySurv** can perform variance estimation through a weighted bootstrap procedure. Package **frailtySurv** is available from the Comprehensive R Archive Network (CRAN) at https://CRAN.R-project.org/package=frailtySurv.

While some of the packages in Table 1 contain synthetic and/or real-world survival datasets, none of them contain functions to simulate clustered data. There exist several other packages capable of simulating survival data, such as the `rmultime` function in the **MST** package (Calhoun, Su, Nunn, and Fan 2018), the `genSurv` function in **survMisc** (Dardis 2016), and **survsim** (Moriña and Navarro 2014), an R package dedicated to simulating survival data. These functions simulate only several frailty distributions. **frailtySurv** contains a rich set of simulation functions, described in Section 2, capable of generating clustered survival data under a wide variety of conditions. The simulation functions in **frailtySurv** are used to empirically verify the implementation of unbiased (bug-free) estimators through several simulated experiments.

The rest of this paper is organized as follows. Sections 2 and 3 describe the data generation and model estimation functions of **frailtySurv**, respectively. Section 4 demonstrates simulation capabilities and results. Section 5 is a case study of two publicly available datasets, including high-risk patients from the Diabetic Retinopathy Study and a large hard drive failure dataset. Finally, Section 6 concludes the paper. The currently supported frailty distributions are described in Appendix A, full simulation results are presented in Appendix B, and Appendix C contains an empirical analysis of runtime and accuracy.

## 2. Data generation

The `genfrail` function in **frailtySurv** can generate clustered survival times under a wide variety of conditions. The survival function at time $t$ of the $j$th observation of cluster $i$, given time-independent covariate $\mathbf{Z}_{ij}$ and frailty variate $\omega_i$, is given by

$$S_{ij}(t|\mathbf{Z}_{ij}, \omega_i) = \exp\left\{-\Lambda_0(t)\,\omega_i e^{\beta^\top \mathbf{Z}_{ij}}\right\}, \qquad (2)$$

where $\Lambda_0(t) = \int_0^t \lambda_0(u)\,du$ is the unspecified cumulative baseline hazard function. In the following sections we describe in detail the various options for setting each component of the above conditional survival function.

### 2.1. Covariates

Covariates can be sampled marginally from normal, uniform, or discrete uniform distributions, as specified by the `covar.distr` parameter. The value of $\beta$ is specified to `genfrail` through the `covar.param` parameter. User-supplied covariates can also be passed as the `covar.matrix` parameter. There is no limit to the covariates' vector dimension. However, the estimation procedure requires the number of clusters to be much higher than the number of covariates. These options are demonstrated in Section 2.7.

### 2.2. Baseline hazard

There are three ways the baseline hazard can be specified to generate survival data: as the inverse cumulative baseline hazard $\Lambda_0^{-1}$, the cumulative baseline hazard $\Lambda_0$, or the baseline hazard $\lambda_0$. If the cumulative baseline hazard function can be directly inverted, then failure times can be computed by

$$T_{ij}^o = \Lambda_0^{-1}\left\{-\ln(U_{ij})\,e^{-\beta^\top \mathbf{Z}_{ij}}/\omega_i\right\}, \qquad (3)$$

where $U_{ij} \sim U(0,1)$ and $T_{ij}^o$ is the failure time of member $j$ of cluster $i$. Consequently, if $\Lambda_0^{-1}$ is provided as parameter `Lambda_0_inv` in `genfrail`, then survival times are determined by Equation 3. This is the most efficient way to generate survival data.

When $\Lambda_0$ cannot be inverted, one can use a univariate root-finding algorithm to solve

$$S_{ij}\left(T_{ij}^o | \mathbf{Z}_{ij}, \omega_i\right) - U_{ij} = 0 \qquad (4)$$

for failure time $T_{ij}^o$. Alternatively, taking the logarithm and solving

$$-\Lambda_0\left(T_{ij}^o\right)\omega_i e^{\beta^\top \mathbf{Z}_{ij}} - \ln U_{ij} = 0 \qquad (5)$$

yields greater numerical stability. Therefore, `genfrail` uses Equation 5 when $\Lambda_0$ is provided as parameter `Lambda_0` in `genfrail` and uses the R function `uniroot`, which is based on Brent's algorithm (Brent 2013).

If neither $\Lambda_0^{-1}$ or $\Lambda_0$ are provided to `genfrail`, then the baseline hazard function $\lambda_0$ must be passed as parameter `lambda_0`. In this case,

$$\Lambda_0(t) = \int_0^t \lambda_0(s)\, ds \qquad (6)$$

is evaluated numerically. Using the `integrate` function in the **stats** package (R Core Team 2018), which implements adaptive quadrature, Equation 5 can be numerically solved for $T_{ij}^o$. This approach is the most computationally expensive since it requires numerical integration to be performed for each observation $ij$ and at each iteration in the root-finding algorithm.

Section 2.7 demonstrates generating data using each of the above methods, which all generate failure times in the range $[0, \infty)$. The computational complexity of each method is $O(n)$ under the assumption that a constant amount of work needs to be performed for each observation. Despite this, the constant amount of work per observation varies greatly depending on how the baseline hazard is specified. Using the inverse cumulative baseline hazard, there exists an analytic solution for each observation and only arithmetic operations are required. Specifying the cumulative baseline hazard requires root finding for each observation, and specifying the baseline hazard requires both root finding and numerical integration for each observation. Since the time to perform root finding and numerical integration is not a function of $n$, the complexity remains linear in each case. Appendix C.1 contains benchmark simulations that compare the timings of each method.

### 2.3. Shared frailty

Shared frailty variates $\omega_1, \ldots, \omega_n$ are generated according to the specified frailty distribution, through parameters `frailty` and `theta` of `genfrail`, respectively. The available distributions are gamma with mean 1 and variance $\theta$; PVF with mean 1 and variance $1 - \theta$; log-normal with mean $\exp(\theta/2)$ and variance $\exp(2\theta) - \exp(\theta)$; and inverse Gaussian with mean 1 and variance $\theta$. `genfrail` can also generate frailty variates from a positive stable (PS) distribution, although estimation is not supported due to the PS having an infinite mean. The supported frailty distributions are described in detail in Appendix A. Specifying parameters that induce a degenerate frailty distribution, or passing `frailty = "none"`, will generate non-clustered data. Hierarchical clustering is currently not supported by **frailtySurv**.

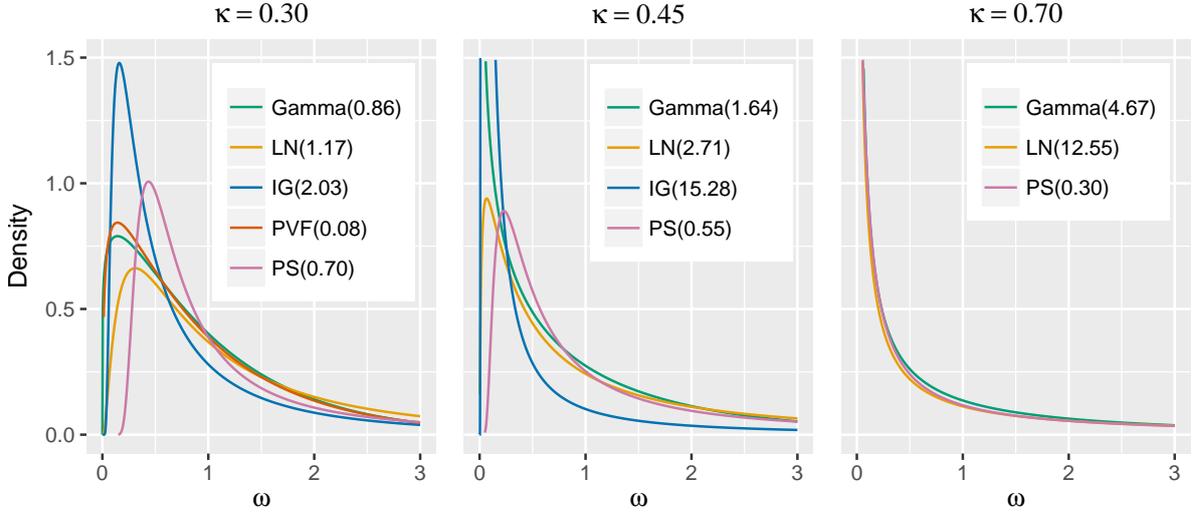

Figure 1: Frailty distribution densities.

The dependence between two cluster members can be measured by Kendall's tau[1], given by

$$\kappa = 4 \int_0^\infty s\mathcal{L}(s)\mathcal{L}^{(2)}(s)\,ds - 1, \qquad (7)$$

where $\mathcal{L}$ is the Laplace transform of the frailty distribution and $\mathcal{L}^{(m)}$, $m = 1, 2, \ldots$ are the $m$th derivatives of $\mathcal{L}$. If failure times of the two cluster members are independent, $\kappa = 0$. Figure 1 shows the densities of the supported distributions for various values of $\kappa$. Note that gamma, IG, and PS are special cases of the PVF. For gamma, LN, and IG, $\kappa = 0$ when $\theta = 0$, and for PVF, $\kappa = 0$ when $\theta = 1$. Also, for gamma and LN, $\lim_{\theta \to \infty} \kappa = 1$, for IG $\lim_{\theta \to \infty} \kappa = 1/2$, for PVF $\lim_{\theta \to 0} \kappa = 1/3$, and for PS $\kappa = 1 - \theta$.

### 2.4. Cluster sizes

In practice, the cluster sizes $m_i$, $i = 1, \ldots, n$, can be fixed or may vary. For example, in the Diabetic Retinopathy Study, two failure times are observed for each subject, corresponding to the left and right eye. Hence, observations are clustered by subject, and each cluster has exactly two members. If instead the observations were clustered by geographical location, the cluster sizes would vary, e.g., according to a discrete power law distribution. `genfrail` is able to generate data with fixed or varying cluster sizes.

For fixed cluster size, the cluster size parameter `K` of `genfrail` is simply an integer. Alternatively, the cluster sizes may be specified by passing a length-N vector of the desired cluster sizes to parameter `K`. To generate varied cluster sizes, `K` is the name of the distribution to generate from, and `K.param` specifies the distribution parameters.

Cluster sizes can be generated from a $k$-truncated Poisson with `K = "poisson"` (Geyer 2018). The truncated Poisson is used to ensure there are no zero-sized clusters and to enforce a

---
[1] Kendall's tau is denoted by $\kappa$ to avoid confusion with $\tau$, the end of the follow-up period.

minimum cluster size. The expected cluster size is given by

$$\begin{cases} \lambda \left(1 - e^{-\lambda} \sum_{j=0}^{k} \frac{\lambda^j}{j!}\right)^{-1} & k = 0, \\ \frac{\lambda - e^{-\lambda} \sum_{j=1}^{k} \frac{\lambda^j}{(j-1)!}}{1 - e^{-\lambda} \sum_{j=0}^{k} \frac{\lambda^j}{j!}} & k > 0, \end{cases} \quad (8)$$

where $\lambda$ is a shape parameter and $k$ is the truncation point such that $\min\{m_1, \ldots, m_n\} > k$. The typical case is with $k = 0$ for a zero-truncated Poisson. For example, with $\lambda = 2$ and $k = 0$, the expected cluster size equals 2.313. The parameters of the $k$-truncated Poisson are determined in `K.param = c(lambda, k)` of `genfrail`.

A discrete Pareto (or zeta) distribution can also be used to generate cluster sizes with `K = "pareto"`. Accurately fitting and generating from a discrete power-law distribution is generally difficult, and `genfrail` uses a truncated discrete Pareto to avoid some of the pitfalls as described in Clauset, Shalizi, and Newman (2009). The probability mass function is given by

$$\mathsf{P}(M = m) = \frac{(m-l)^{-s}/\zeta(s)}{\sum_{j=1}^{u-l} j^{-s}/\zeta(s)}, \quad s > 1, \ u > l, \ m = l+1, \ldots, u, \quad (9)$$

where $\zeta(s)$ is the Riemann zeta function, $s$ is a scaling parameter, $l$ is the noninclusive lower bound, and $u$ is the inclusive upper bound. With large enough $u$ and $s \gg 1$, the distribution behaves similar to the discrete Pareto distribution and the expected cluster size equals $\frac{1}{\zeta(s)} \sum_{j=1}^{\infty} \frac{1}{j^{s-1}}$. The distribution parameters are specified as `K.param = c(s, u, l)`. Finally, a discrete uniform distribution can be specified by `K = "uniform"` in `genfrail`. The respective parameters to `K.param` are `c(l, u)`, where $l$ is the noninclusive lower bound and $u$ is the inclusive upper bound. Similar to the truncated zeta, the support is $\{l+1, \ldots, u\}$ while each cluster size is uniformly selected from this set of values. Since the lower bound is noninclusive, the expected cluster size equals $(1 + l + u)/2$.

### 2.5. Censoring

The observed times $T_{ij}$ and failure indicators $\delta_{ij}$ are determined by the failure times $T_{ij}^o$ and right-censoring times $C_{ij}$ such that the observed time of observation $ij$ is given by

$$T_{ij} = \min\left(T_{ij}^o, C_{ij}\right), \quad j = 1, \ldots, m_i \ \ i = 1, \ldots, n, \quad (10)$$

and the failure indicator is given by

$$\delta_{ij} = I\left(T_{ij}^o \leq C_{ij}\right), \quad j = 1, \ldots, m_i \ \ i = 1, \ldots, n. \quad (11)$$

Currently, only right-censoring is supported by **frailtySurv**. The censoring distribution is specified by the parameters `censor.distr` and `censor.param` for the distribution name and parameters' vector, respectively. A normal distribution is used by default. A log-normal censoring distribution is specified by `censor.distr = "lognormal"` and `censor.param = c(mu, sigma)`, where `mu` is the mean and `sigma` is the standard deviation of the censoring distribution. Lastly, a uniform censoring distribution can be specified by `censor.distr = "uniform"` and `censor.param = c(lower, upper)` for the lower and upper bounds on the interval, respectively.

Sometimes a particular censoring rate is desired. Typically, the censoring distribution parameters are varied to obtain a desired censoring rate. `genfrail` can avoid this effort on behalf of the user by letting the desired censoring rate be specified instead. In this case, the appropriate parameters for the censoring distribution are determined to achieve the desired censoring rate, given the generated failure times.

Let $F$ and $G$ be the failure time and censoring time cumulative distributions, respectively. Then, the censoring rate equals

$$\mathsf{E}\left\{I(T_{11}^o > C_{11})\right\} = \int_0^\infty G(t)dF(t) , \tag{12}$$

where the expectation of $I(T_{11}^o > C_{11})$ equals the expectation of any random subject from the population. The above formula can be estimated by

$$\widehat{\mathsf{E}}\left\{I(T_{11}^o > C_{11})\right\} = \int_0^\infty G(t)d\widehat{F}(t), \tag{13}$$

where $\widehat{F}$ is the empirical cumulative distribution function. To obtain a particular censoring rate $0 < R < 1$, as a function of the parameters of $G$, one can solve

$$R - \widehat{\mathsf{E}}\left\{I(T_{11}^o > C_{11})\right\} = 0 . \tag{14}$$

For example, if $G$ is the normal cumulative distribution function with mean $\mu$ and variance $\sigma^2$, $\sigma^2$ should be pre-specified (otherwise the problem is non-identifiable), and Equation 14 is solved for $\mu$. This method works with any empirical distribution of failure times. `genfrail` uses this approach to achieve a desired censoring rate, specified by `censor.rate`, with normal, log-normal, or uniform censoring distributions. Lastly, user-supplied censorship times can be supplied through the `censor.time` parameter, which must be a vector of length N * K, where N is the number of clusters and K is the size of each cluster. Because of this, `censor.time` cannot be used with variable-sized clusters.

### 2.6. Rounding

In some applications the observed times are rounded and tied failure times can occur. For example, the age at onset of certain diseases are often recorded as years of age rounded to the nearest integer. To simulate tied data, the simulated observed times may optionally be rounded to the nearest integer of multiple of $B$ by

$$\dot{T}_{ij} = B\left\lfloor \frac{T_{ij}}{B} + 0.5 \right\rfloor . \tag{15}$$

If $B = 1$, the observed times are simply rounded to the nearest integer. The value of $B$ is specified by the parameter `round.base` of `genfrail`, with the default being the non-rounded setting.

### 2.7. Examples

The best way to see how `genfrail` works is through examples. R and **frailtySurv** versions are given by the following commands.

```
R> R.Version()$version.string
```

```
[1] "R version 3.4.3 (2017-11-30)"
```

```
R> packageDescription("frailtySurv", fields = "Version")
```

```
[1] "1.3.5"
```

Consider the survival model defined in Equation 2 with baseline hazard function

$$\lambda_0\left(t\right) = \left\{d\left(ct\right)^d\right\} t^{-1}, \tag{16}$$

where $c = 0.01$ and $d = 4.6$. Let Gamma $(2)$ be the frailty distribution, two independent standard normally distributed covariates, and $N\left(130, 15^2\right)$ the censoring distribution. The resulting survival times are representative of a late onset disease and with $\sim 40\%$ censoring rate. Generating survival data from this model, with 300 clusters and 2 members within each cluster, is accomplished by[2]

```
R> set.seed(2015)
R> dat <- genfrail(N = 300, K = 2, beta = c(log(2), log(3)),
+    frailty = "gamma", theta = 2,
+    lambda_0 = function(t, c = 0.01, d = 4.6) (d * (c * t) ^ d) / t)
R> head(dat, 3)

  family rep      time status        Z1        Z2
1      1   1  87.95447      1 -1.5454484  0.9944159
2      1   2 110.04615      0 -0.5283932 -0.9053164
3      2   1 119.94127      1 -1.0867588  0.5240979
```

Similarly, to generate survival data with uniform covariates from, e.g., 0.1 to 0.2, specify `covar.distr = "uniform"` and `covar.param = c(0.1, 0.2)` in the above example. The covariates may also be specified explicitly in a `c(N * K, length(beta))` matrix as the `covar.matrix` parameter.

In the above example, the baseline hazard function was specified by the `lambda_0` parameter. The same dataset can be generated more efficiently using the `Lambda_0` parameter if the cumulative baseline hazard function is known. This is accomplished by integrating Equation 16 to get the cumulative baseline hazard function

$$\Lambda_0\left(t\right) = \left(ct\right)^d \tag{17}$$

and passing this function as an argument to `Lambda_0` when calling `genfrail`:

```
R> set.seed(2015)
R> dat.cbh <- genfrail(N = 300, K = 2, beta = c(log(2),log(3)),
+    frailty = "gamma", theta = 2,
+    Lambda_0 = function(t, c = 0.01, d = 4.6) (c * t) ^ d)
R> head(dat.cbh, 3)
```

---

[2] Note that `N` and `K` are the parameters of `genfrail` that correspond to math notation $n$ (number of clusters) and $m_i$ (cluster size), respectively.

```
  family rep       time status          Z1         Z2
1      1   1   87.95447      1  -1.5454484  0.9944159
2      1   2  110.04615      0  -0.5283932 -0.9053164
3      2   1  119.94127      1  -1.0867588  0.5240979
```

The cumulative baseline hazard in Equation 17 is invertible and it would be even more efficient to specify $\Lambda_0^{-1}$ as

$$\Lambda_0^{-1}(t) = c^{-1} t^{1/d} . \tag{18}$$

This avoids the numerical integration, required by Equation 6, and root finding, required by Equation 5. Equation 18 should be passed to `genfrail` as the `Lambda_0_inv` parameter, again producing the same data when the same seed is used:

```
R> set.seed(2015)
R> dat.inv <- genfrail(N = 300, K = 2, beta = c(log(2),log(3)),
+    frailty = "gamma", theta = 2,
+    Lambda_0_inv = function(t, c = 0.01, d = 4.6) (t ^ (1 / d)) / c)
R> head(dat.inv, 3)

  family rep       time status          Z1         Z2
1      1   1   87.95449      1  -1.5454484  0.9944159
2      1   2  110.04615      0  -0.5283932 -0.9053164
3      2   1  119.94127      1  -1.0867588  0.5240979
```

A different frailty distribution can be specified while ensuring an expected censoring rate by using the `censor.rate` parameter. For example, consider a PVF (0.3) frailty distribution while maintaining the 40% censoring rate in the previous example. The censoring distribution parameters are determined by `genfrail` as described in Section 2.5 by specifying `censor.rate = 0.4`. This avoids the need to manually adjust the censoring distribution to achieve a particular censoring rate. The respective code and output are:

```
R> set.seed(2015)
R> dat.pvf <- genfrail(N = 300, K = 2, beta = c(log(2),log(3)),
+    frailty = "pvf", theta = 0.3, censor.rate = 0.4,
+    Lambda_0_inv = function(t, c = 0.01, d = 4.6) (t ^ (1 / d)) / c)
R> summary(dat.pvf)

genfrail created     : 2018-06-14 13:46:36
Observations         : 600
Clusters             : 300
Avg. cluster size    : 2.00
Right censoring rate : 0.39
Covariates           : normal(0, 1)
Coefficients         : 0.6931, 1.0986
Frailty              : pvf(0.3)
Baseline hazard      : Lambda_0
                       = function (t, tau = 4.6, C = 0.01) (t^(1/tau))/C
```

# 3. Model estimation

The `fitfrail` function in **frailtySurv** estimates the regression coefficient vector $\beta$, the frailty distribution's parameter $\theta$, and the non-parametric cumulative baseline hazard $\Lambda_0$. The observed data consist of $\{T_{ij}, \mathbf{Z}_{ij}, \delta_{ij}\}$ for $i = 1, \ldots, n$ and $j = 1, \ldots, m_i$, where the $n$ clusters are independent. `fitfrail` takes a complete observation approach, and observations with missing values are ignored with a warning.

There are two estimation strategies that can be used. The log-likelihood can be maximized directly, by using control parameter `fitmethod = "loglik"`, or a system of score equations can be solved with control parameter `fitmethod = "score"`. Both methods have comparable computational requirements and yield comparable results. In both methods, the estimation procedure consists of a doubly-nested loop, with an outer loop that evaluates the objective function and gradients and an inner loop that estimates the piecewise constant hazard, performing numerical integration at each time step if necessary. As a result, the estimator implemented in **frailtySurv** has computationally complexity on the order of $O(n^2)$.

## 3.1. Log-likelihood

The full likelihood can be written as

$$
\begin{aligned}
L(\beta, \theta, \Lambda_0) &= \prod_{i=1}^{n} \int \prod_{j=1}^{m_i} \{\lambda_{ij}(T_{ij}|\mathbf{Z}_{ij}, \omega)\}^{\delta_{ij}} S_{ij}(T_{ij}|\mathbf{Z}_{ij}, \omega) f(\omega) d\omega \\
&= \prod_{i=1}^{n} \prod_{j=1}^{m_i} \left\{\lambda_0(T_{ij}) e^{\beta^\top \mathbf{Z}_{ij}}\right\}^{\delta_{ij}} \prod_{i=1}^{n} (-1)^{N_{i.}(\tau)} \mathcal{L}^{(N_{i.}(\tau))} \{H_{i.}(\tau)\},
\end{aligned} \quad (19)
$$

where $\tau$ is the end of follow-up period, $f$ is the frailty's density function, $N_{ij}(t) = \delta_{ij} I(T_{ij} \leq t)$, $N_{i.}(t) = \sum_{j=1}^{m_i} N_{ij}(t)$, $H_{ij}(t) = \Lambda_0(T_{ij} \wedge t) e^{\beta^\top \mathbf{Z}_{ij}}$, and $H_{i.}(t) = \sum_{j=1}^{m_i} H_{ij}(t)$, $j = 1, \ldots, m_i$, $i = 1, \ldots, n$. Note that the $m$th derivative of the Laplace transform evaluated at $H_{i.}(\tau)$ equals $(-1)^{N_{i.}(\tau)} \int \omega^{N_{i.}(\tau)} \exp\{-\omega H_{i.}(\tau)\} f(\omega) d\omega$, $i = 1, \ldots, n$. The log-likelihood equals

$$
\ell(\beta, \theta, \Lambda_0) = \sum_{i=1}^{n} \sum_{j=1}^{m_i} \delta_{ij} \log\left\{\lambda_0(T_{ij}) e^{\beta^\top \mathbf{Z}_{ij}}\right\} + \sum_{i=1}^{n} \log \mathcal{L}^{\{N_{i.}(\tau)\}} \{H_{i.}(\tau)\}. \quad (20)
$$

Evidently, to obtain estimators $\widehat{\beta}$ and $\widehat{\theta}$ based on the log-likelihood, an estimator of $\Lambda_0$, denoted by $\widehat{\Lambda}_0$, is required. For given values of $\beta$ and $\theta$, $\Lambda_0$ is estimated by a step function with jumps at the ordered observed failure times $\tau_k$, $k = 1, \ldots, K$, defined by

$$
\Delta \widehat{\Lambda}_0(\tau_k) = \frac{d_k}{\sum_{i=1}^{n} \psi_i\left(\gamma, \widehat{\Lambda}_0, \tau_{k-1}\right) \sum_{j=1}^{m_i} Y_{ij}(\tau_k) e^{\beta^\top \mathbf{Z}_{ij}}}, \quad k = 1, \ldots, K, \quad (21)
$$

where $d_k$ is the number of failures at time $\tau_k$, $\psi_i(\gamma, \Lambda, t) = \phi_{2i}(\gamma, \Lambda, t) / \phi_{1i}(\gamma, \Lambda, t)$, $Y_{ij}(t) = I(T_{ij} \geq t)$, and

$$
\phi_{ai}(\gamma, \Lambda_0, t) = \mathcal{L}^{(N_{i.}(t) + a - 1)} \{H_{i.}(t)\} \quad a = 1, 2.
$$

For the detailed derivation of the above baseline hazard estimation the reader is referred to Gorfine *et al.* (2006). The estimator of the cumulative baseline hazard at time $\tau_k$ is given by

$$
\widehat{\Lambda}_0(\tau_k) = \sum_{l=1}^{k} \Delta \widehat{\Lambda}_0(\tau_l), \quad (22)
$$

and is a function of $\sum_{i=1}^{m_i} \widehat{\Lambda}_0 (T_{ij} \wedge \tau_{k-1}) e^{\beta^\top \mathbf{Z}_{ij}}$, i.e., at each $\tau_k$, the cumulative baseline hazard estimator is a function of $\widehat{\Lambda}_0(t)$ with $t < \tau_k$. Then, for obtaining $\widehat{\beta}$ and $\widehat{\theta}$, $\widehat{\Lambda}_0$ is substituted into $\ell(\beta, \theta, \Lambda_0)$.

In summary, the estimation procedure of Gorfine *et al.* (2006) consists of the following steps:

**Step 1.** Use standard Cox regression software to obtain initial estimates of $\beta$, and set the initial value of $\theta$ to be its value under within-cluster independence or under very week dependency (see also the discussion at the end of Section 3.2).

**Step 2.** Use the current values of $\beta$ and $\theta$ to estimate $\Lambda_0$ based on the estimation procedure defined by Equation 21.

**Step 3.** Using the current value of $\widehat{\Lambda}_0$, estimate $\beta$ and $\theta$ by maximizing $l(\beta, \theta, \widehat{\Lambda}_0)$.

**Step 4.** Iterate between Steps 2 and 3 until convergence.

For frailty distributions with no closed-form Laplace transform, the integral can be evaluated numerically. This adds a considerable overhead to each iteration in the estimation procedure since the integrations must be performed for the baseline hazard estimator that is required for estimating $\beta$ and $\theta$, as $H_{i\cdot}(\tau) = \sum_{i=1}^{m_i} \Lambda_0(T_{ij} \wedge \tau) e^{\beta^\top \mathbf{Z}_{ij}}$.

With control parameter `fitmethod = "loglik"`, the log-likelihood is the objective function maximized directly with respect to $\gamma = (\beta^\top, \theta)^\top$, for any given $\Lambda_0$, by `optim` in the **stats** package using the L-BFGS-B algorithm (Byrd, Lu, Nocedal, and Zhu 1995). Box constraints specify bounds on the frailty distribution parameters, typically $\theta \in (0, \infty)$ except for PVF which has $\theta \in (0, 1)$. Convergence is determined by the relative reduction in the objective function through the `reltol` control parameter. By default, this is $10^{-6}$.

As an example, consider fitting a model to the data generated in Section 2. The following result shows that convergence is reached after 11 iterations and 15.8 seconds, running Red Hat 6.5, R version 3.2.2, and 2.6 GHz Intel Sandy Bridge processor:

```
R> fit <- fitfrail(Surv(time, status) ~ Z1 + Z2 + cluster(family),
+    dat, frailty = "gamma", fitmethod = "loglik")
R> fit

Call: fitfrail(formula = Surv(time, status) ~ Z1 + Z2 + cluster(family),
    dat = dat, frailty = "gamma", fitmethod = "loglik")

    Covariate      Coefficient
           Z1            0.719
           Z2            1.194

Frailty distribution   gamma(1.716), VAR of frailty variates = 1.716
Log-likelihood         -2507.725
Converged (method)     11 iterations, 6.75 secs (maximized log-likelihood)
```

### 3.2. Score equations

Instead of maximizing the log-likelihood, one can solve the score equations. The score function with respect to $\beta$ is given by

$$\begin{aligned}
\mathbf{U}_\beta = \frac{\partial}{\partial \beta}\ell(\beta,\theta,\Lambda_0) &= \sum_{i=1}^{n}\left[\sum_{j=1}^{m_i}\delta_{ij}\mathbf{Z}_{ij} + \frac{\frac{\partial}{\partial\beta}H_{i.}(\tau)\frac{\partial}{\partial H_{i.}(\tau)}\mathcal{L}^{\{N_{i.}(\tau)\}}(H_{i.}(\tau))}{\mathcal{L}^{\{N_{i.}(\tau)\}}(H_{i.}(\tau))}\right] \\
&= \sum_{i=1}^{n}\left[\sum_{j=1}^{m_i}\delta_{ij}\mathbf{Z}_{ij} + \sum_{j=1}^{m_i}H_{ij}(T_{ij})\mathbf{Z}_{ij}\frac{\mathcal{L}^{\{N_{i.}(\tau)+1\}}(H_{i.}(\tau))}{\mathcal{L}^{\{N_{i.}(\tau)\}}(H_{i.}(\tau))}\right] . \quad (23)
\end{aligned}$$

Note that $\mathcal{L}^{(N_{i.}(\tau)+1)}\{H_{i.}(\tau)\}/\mathcal{L}^{(N_{i.}(\tau))}\{H_{i.}(\tau)\}$ corresponds to $\psi_i$ in Gorfine *et al.* (2006). The score function with respect to $\theta$ is given by

$$\mathbf{U}_\theta = \frac{\partial}{\partial\theta}\ell(\beta,\theta,\Lambda_0) = \sum_{i=1}^{n}\frac{\frac{\partial}{\partial\theta}\mathcal{L}^{(N_{i.}(\tau))}(H_{i.}(\tau))}{\mathcal{L}^{(N_{i.}(\tau))}(H_{i.}(\tau))} . \quad (24)$$

The score equations are given by $\mathbf{U}(\beta,\theta,\Lambda_0) = (\mathbf{U}_\beta, \mathbf{U}_\theta) = \mathbf{0}$ and the estimator of $\gamma = (\beta^\top, \theta)$ is defined as the value of $(\beta^\top, \theta)$ that solves the score equations for any given $\Lambda_0$. Specifically, the only change required in the above summary of the estimation procedure, is to replace Step 3 with the following

**Step 3'.** Using the current value of $\widehat{\Lambda}_0$, estimate $\beta$ and $\theta$ by solving $\mathbf{U}(\beta,\theta,\widehat{\Lambda}_0) = \mathbf{0}$.

**frailtySurv** uses Newton's method implemented by the **nleqslv** package to solve the system of equations (Hasselman 2017). Convergence is reached when the relative reduction of each parameter estimate or absolute value of each normalized score is below the threshold specified by `reltol` or `abstol`, respectively. The default is a relative reduction of $\widehat{\gamma}$ less than $10^{-6}$, i.e., `reltol = 1e-6`.

As an example, in the following lines of code and output we consider again the data generated in Section 2. The results are comparable to the fitted model in Section 3.1. The score equations can usually be solved in fewer iterations than maximizing the likelihood, although solving the system of equations requires more work in each iteration. For this reason, maximizing the likelihood is typically more computationally efficient for large datasets when a permissive convergence criterion is specified.

```
R> fit.score <- fitfrail(Surv(time, status) ~ Z1 + Z2 + cluster(family),
+    dat, frailty = "gamma", fitmethod = "score")
R> fit.score

Call: fitfrail(formula = Surv(time, status) ~ Z1 + Z2 + cluster(family),
    dat = dat, frailty = "gamma", fitmethod = "score")

    Covariate    Coefficient
           Z1          0.719
           Z2          1.194
```

```
Frailty distribution    gamma(1.716), VAR of frailty variates = 1.716
Log-likelihood          -2507.725
Converged (method)      10 iterations, 6.50 secs (solved score equations)
```

L-BFGS-B, used for maximizing the log-likelihood, allows for (possibly open-ended) box constraints. In contrast, Newton's method, used for solving the system of score equations, does not support the use of box constraints and, therefore, has a risk of converging to a degenerate parameter value. In this case, it is more important to have a sensible starting value. In both estimation methods, the regression coefficient vector $\beta$ is initialized to the estimates given by `coxph` with no shared frailty. The frailty distribution parameters are initialized such that the dependence between members in each cluster is small, i.e, with $\kappa \approx 0.3$.

### 3.3. Baseline hazard

The estimated cumulative baseline hazard defined by Equation 22 is accessible from the resulting model object through the `fit$Lambda` member, which provides a `data.frame` with the estimates at each observed failure time, or the `fit$Lambda.fun` member, which defines a scalar R function that takes a time argument and returns the estimated cumulative baseline hazard. The estimated survival curve or cumulative baseline hazard can also be summarized by the `summary` method for objects returned by `fitfrail` resulting in a `data.frame`. In the example below, the `n.risk` column contains the number of observations still at risk at time $t-$ and the `n.event` column contains the number of failures from the previous time listed to time $t+$. The output is similar to that of the `summary` method for 'survfit' objects in the **survival** package.

```
R> head(summary(fit), 3)

      time n.risk n.event      surv
1 23.37616    600       1 0.9992506
2 24.38503    599       1 0.9984604
3 25.14435    598       1 0.9976600

R> tail(summary(fit), 3)

        time n.risk n.event        surv
384 139.5629     42       1 0.0016570493
385 140.5862     39       1 0.0011509892
386 141.3295     36       1 0.0007665802
```

By default, the survival curve estimates at observed failure times are returned. Estimates at the censored observed times are included if `censored = TRUE` is passed to the `summary` method for 'fitfrail' objects. The cumulative baseline hazard estimates are summarized by parameter `type = "cumhaz"`. The estimates can also be evaluated at specific times passed to the `summary` method for 'fitfrail' objects through the `Lambda.times` parameter, demonstrated by:

```
R> summary(fit, type = "cumhaz", Lambda.times = c(20, 50, 80, 110))

  time n.risk n.event     cumhaz
1   20    600       0 0.00000000
2   50    566      34 0.03248626
3   80    439     127 0.33826069
4  110    274     147 1.69720757
```

### 3.4. Standard errors

There are two ways the standard errors can be obtained for a fitted model. The covariance matrix of $\widehat{\gamma}$, the estimators of the regression coefficients and the frailty parameter, can be obtained explicitly based on the sandwich-type consistent estimator described in Gorfine *et al.* (2006) and Zucker *et al.* (2008). The covariance matrix is calculated by the vcov function applied to the 'fitfrail' object returned by fitfrail. Optionally, standard errors can also be obtained in the call to fitfrail by passing se = TRUE. Using the above fitted model, the covariance matrix of $\widehat{\gamma}$ is obtained by

```
R> COV.est <- vcov(fit)
R> sqrt(diag(COV.est))

        Z1         Z2    theta.1
0.09343685 0.12673624 0.36020143
```

**frailtySurv** can also estimate standard errors through a weighted bootstrap approach, in which the variance of both $\widehat{\gamma}$ and $\widehat{\Lambda}_0$ are determined[3]. The weighted bootstrap procedure consists of independent and identically distributed positive random weights applied to each cluster. This is in contrast to a nonparametric bootstrap, wherein each bootstrap sample consists of a random sample of clusters with replacement. The resampling procedure of the nonparametric bootstrap usually yields an increased number of ties compared to the original data, which sometimes causes convergence problems. Therefore, we adopt the weighted bootstrap approach which does not change the number of tied observations in the original data. The weighted bootstrap is summarized as follows.

1. Sample $n$ random values $\{v_i^*, i = 1, \ldots, n\}$ from an exponential distribution with mean 1. Standardize the values by the empirical mean to obtain standardized weights $v_1, \ldots, v_n$.

2. In the estimation procedure, each function of the form $\sum_{i=1}^n h(\mathbf{T}_i, \delta_i, \mathbf{Z}_i)$ is replaced be the corresponding weighted function $\sum_{i=1}^n v_i h(\mathbf{T}_i, \delta_i, \mathbf{Z}_i)$, where $\mathbf{T}_i = (T_{i1}, \ldots, T_{im_i})$, $\delta_i = (\delta_{i1}, \ldots, \delta_{im_i})$, and $\mathbf{Z}_i = (Z_{i1}, \ldots, Z_{im_i})$, $i = 1, \ldots, n$.

3. Repeat Steps 1–2 $B$ times and take the empirical variance (and covariance) of the $B$ parameter estimates to obtain the weighted bootstrap variance (and covariance).

For smaller datasets, this process is generally more time-consuming than the explicit estimator. If the **parallel** package is available, all available cores are used to obtain the bootstrap parameter estimates in parallel (R Core Team 2018). Without the **parallel** package, vcov runs in serial.

---

[3]The sandwich estimator currently only provides the covariance matrix of $\widehat{\gamma}$ and not $\widehat{\Lambda}_0$.

```
R> set.seed(2015)
R> COV.boot <- vcov(fit, boot = TRUE, B = 500)
R> sqrt(diag(COV.boot))[1:8]

               Z1               Z2          theta.1 Lambda.  0.00000
      0.0742560635     0.0984509739     0.2568936409     0.0000000000
Lambda. 23.37616 Lambda. 24.38503 Lambda. 25.14435 Lambda. 25.33731
      0.0006340182     0.0010267995     0.0012781985     0.0014768459
```

In the preceding example, the full covariance matrix for $\left(\widehat{\gamma}, \widehat{\Lambda}_0\right)$ is obtained. If only certain time points of the estimated cumulative baseline hazard function are desired, these can be specified by the `Lambda.times` parameter. Since calls to the `vcov` method for 'fitfrail' objects are typically computationally expensive, the results are cached when the same arguments are provided.

### 3.5. Control parameters

Control parameters provided to `fitfrail` determine the speed, accuracy, and type of estimates returned. The default `control` parameters to `fitfrail` are given by calling the function `fitfrail.control()`. This returns a named list with the following members.

**fitmethod:** Parameter estimation procedure. Either `"score"` to solve the system of score equations or `"loglik"` to estimate using the log-likelihood. Default is `"loglik"`.

**abstol:** Absolute tolerance for convergence. Default is 0 (ignored).

**reltol:** Relative tolerance for convergence. Default is `1e-6`.

**maxit:** The maximum number of iterations before terminating the estimation procedure. Default is 100.

**int.abstol:** Absolute tolerance for numerical integration convergence. Default is 0 (ignored).

**int.reltol:** Relative tolerance for numerical integration convergence. Default is 1.

**int.maxit:** The maximum number of function evaluations in numerical integration. Default is 1000.

**verbose:** If `verbose = TRUE`, the parameter estimates and log-likelihood are printed at each iteration. Default is `FALSE`.

The parameters `int.abstol`, `int.reltol`, and `int.maxit` are only used for frailty distributions that require numerical integration, as they specify convergence criteria of numerical integration in the estimation procedure inner loop. These control parameters can be adjusted to obtain an speed-accuracy tradeoff, whereby lower `int.abstol` and `int.reltol` (and higher `int.maxit`) yield more accurate numerical integration at the expense of more work performed in the inner loop of the estimation procedure.

The `abstol`, `reltol`, and `maxit` parameters specify convergence criteria of the outer loop of the estimation procedure. Similar to the numerical integration convergence parameters, these can also be adjusted to obtain a speed-accuracy tradeoff using either estimation procedure (`fitmethod = "loglik"` or `fitmethod = "score"`). If `fitmethod = "loglik"`, convergence is reached when the absolute or relative reduction in log-likelihood is less than `abstol` or `reltol`, respectively. Using `fitmethod = "score"` and specifying `abstol > 0` (with `reltol = 0`), convergence is reached when the absolute value of each score equation is below `abstol`. Alternatively, using `fitmethod = "score"` and specifying `reltol > 0` (with `abstol = 0`), convergence is reached when the relative reduction of parameter estimates $\widehat{\gamma}$ is below `reltol`. Note that with `fitmethod = "score"`, `abstol` and `reltol` correspond to parameters `ftol` and `xtol` of **nleqslv**::`nleqslv`, respectively. The default convergence criteria were chosen to yield approximately the same results with either estimation strategy.

### 3.6. Model object

The resulting model object returned by `fitfrail` contains the regression coefficients' vector, the frailty distribution's parameters, and the cumulative baseline hazard. Specifically:

`beta:` Estimated regression coefficients' vector named by the input data columns.

`theta:` Estimated frailty distribution parameter.

`loglik:` The resulting log-likelihood.

`Lambda:` `data.frame` with the cumulative baseline hazard at the observed failure times.

`Lambda.all:` `data.frame` with the cumulative baseline hazard at all observed times.

`Lambda.fun:` Scalar R function that returns the cumulative baseline hazard at any time point.

The model object also contains some standard attributes, such as `call` for the function call. If `se = TRUE` was passed to `fitfrail`, then the model object will also contain members `se.beta` and `se.theta` for the standard error of the regression coefficients' vector and frailty parameter estimates, respectively.

## 4. Simulation

As an empirical proof of implementation, and to demonstrate flexibility, several simulations were conducted. The `simfrail` function can be used to run a variety of simulation settings. Simulations are run in parallel if the **parallel** package is available, and the `mc.cores` parameter specifies how many processor cores to use. For example,

```
R> set.seed(2015)
R> sim <- simfrail(1000,
+    genfrail.args = alist(beta = c(log(2),log(3)), frailty = "gamma",
+      censor.rate = 0.30, N = 300, K = 2, theta = 2,
+      covar.distr = "uniform", covar.param = c(0, 1),
+      Lambda_0 = function(t, c = 0.01, d = 4.6) (c * t) ^ d),
+    fitfrail.args = alist(
+      formula = Surv(time, status) ~ Z1 + Z2 + cluster(family),
+      frailty = "gamma"), Lambda.times = 1:120)
R> summary(sim)
```

```
Simulation: 1000 reps, 300 clusters (avg. size 2), gamma frailty
Serial runtime (s): 9680.18 (9.68 +/- 1.53 per rep)
         beta.1 beta.2 theta.1 Lambda.30 Lambda.60 Lambda.90
value    0.6931 1.0986  2.0000  0.003933   0.09539    0.6159
mean.hat 0.6821 1.0929  1.9752  0.003995   0.09716    0.6236
sd.hat   0.2472 0.2529  0.2659  0.001876   0.02248    0.1387
mean.se  0.3130 0.3156  0.3442        NA        NA        NA
cov.95CI 0.9890 0.9850  0.9780        NA        NA        NA
```

The above results indicate that the empirical coverage rates are reasonably close to the nominal 95% coverage rate. These results can also be compared to the estimates obtained by `coxph` which applies the PPL approach with gamma frailty model:

```
R> set.seed(2015)
R> sim.coxph <- simcoxph(1000,
+    genfrail.args = alist(beta = c(log(2), log(3)), frailty = "gamma",
+      censor.rate = 0.30, N = 300, K = 2, theta = 2,
+      covar.distr = "uniform", covar.param = c(0, 1),
+      Lambda_0 = function(t, c = 0.01, d = 4.6) (c * t) ^ d),
+    coxph.args = alist(
+      formula = Surv(time, status) ~ Z1 + Z2 + frailty.gamma(family)),
+    Lambda.times = 1:120)
R> summary(sim.coxph)

Simulation: 1000 reps, 300 clusters (avg. size 2), gamma frailty
Serial runtime (s): 113.27 (0.11 +/- 0.02 per rep)
         beta.1 beta.2 theta.1 Lambda.30 Lambda.60 Lambda.90
value    0.6931 1.0986  2.0000  0.003933   0.09539    0.6159
mean.hat 0.6783 1.0913  1.9843  0.004003   0.09754    0.6282
sd.hat   0.2447 0.2522  0.2665  0.001869   0.02221    0.1375
mean.se  0.2456 0.2468      NA        NA        NA        NA
cov.95CI 0.9470 0.9440      NA        NA        NA        NA
```

The above output indicates that the **frailtySurv** and PPL approach with gamma frailty distribution provide similar results. Note that the `theta.1` mean SE and coverage rate are `NA` since `coxph` does not provide the SE for the estimated frailty distribution parameter.

The correlation between regression coefficient and frailty distribution parameter estimates of both methods is given by

```
R> sapply(names(sim)[grepl("^hat.beta|^hat.theta", names(sim))],
+    function(name) cor(sim[[name]], sim.coxph[[name]]))

 hat.beta.1  hat.beta.2 hat.theta.1
  0.9912442   0.9911590   0.9982390
```

The mean correlation between cumulative baseline hazard estimates is given by

```
R> mean(sapply(names(sim)[grepl("^hat.Lambda", names(sim))],
+    function(name) cor(sim[[name]], sim.coxph[[name]])), na.rm = TRUE)

[1] 0.9867021
```

Full simulation results are provided in Appendix B and include the following settings: gamma frailty with various number of clusters; large cluster size; discrete observed times; oscillating baseline hazard; PVF frailty with fixed and random cluster size; log-normal frailty; and inverse Gaussian frailty. It is evident that for all the available frailty distributions our estimation procedure and implementation work very well in terms of bias, and the sandwich-type variance estimator is dramatically improved as the cluster size increases (for example, from 2 to 6). The bootstrap variance estimators are shown to be accurate even with small cluster size.

# 5. Case study

To demonstrate the applicability of **frailtySurv**, results are obtained for two different datasets. The first is a clinical dataset, for which several benchmark results exist. The second is a hard drive failure dataset from a large cloud backup storage provider. Both datasets are provided with **frailtySurv** as data("drs", package = "frailtySurv") and data("hdfail", package = "frailtySurv"), respectively.

## 5.1. Diabetic Retinopathy Study

The Diabetic Retinopathy Study (DRS) was performed to determine whether the onset of blindness in 197 high-risk diabetic patients could be delayed by laser treatment (The Diabetic Retinopathy Study Research Group 1976). The treatment was administered to one randomly-selected eye in each patient, leaving the other eye untreated. Thus, there are 394 observations which are clustered by patient due to unobserved patient-specific effects. A failure occurred when visual acuity dropped to below 5/200, and approximately 61% of observations are right-censored. All patients had a visual acuity of at least 20/100 at the beginning of the study. A model with gamma shared frailty is estimated from the data.

```
R> data("drs", package = "frailtySurv")
R> fit.drs <- fitfrail(Surv(time, status) ~ treated + cluster(subject_id),
+    drs, frailty = "gamma")
R> COV.drs <- vcov(fit.drs)
R> fit.drs

Call: fitfrail(formula = Surv(time, status) ~ treated + cluster(subject_id),
    dat = drs, frailty = "gamma")

    Covariate       Coefficient
      treated            -0.918

Frailty distribution    gamma(0.876), VAR of frailty variates = 0.876
Log-likelihood          -1005.805
Converged (method)      7 iterations, 1.36 secs (maximized log-likelihood)
```

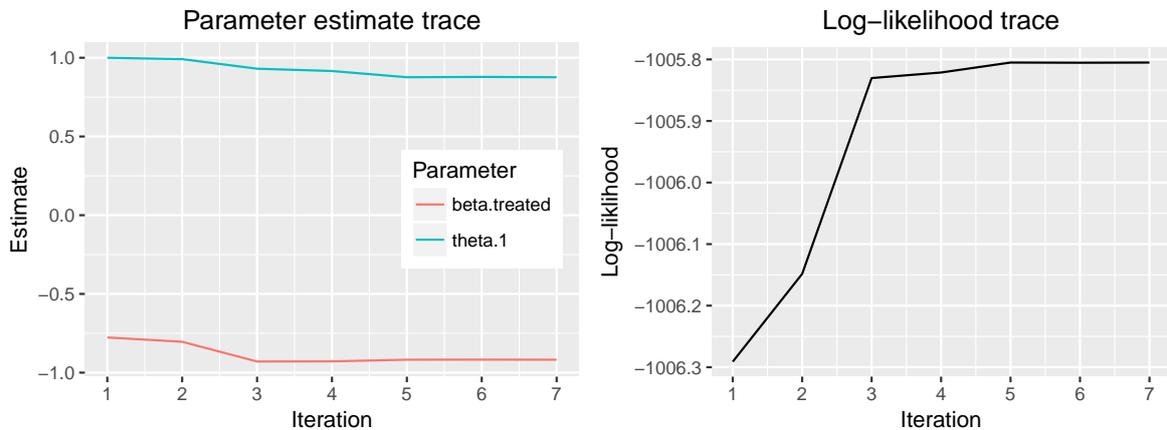

Figure 2: Parameter and log-likelihood trace.

```
R> sqrt(diag(COV.drs))
```

```
   treated    theta.1
0.1975261  0.3782775
```

The regression coefficient for the binary `treated` variable is estimated to be $-0.918$ with $0.198$ estimated standard error, which indicates a 60% decrease in hazard with treatment. The $p$ value for testing the null hypothesis that the treatment has no effect against a two sided alternative equals $3.5 \times 10^{-6}$ (calculated by `2 * pnorm(-0.918/0.198)`). The parameter trace can be plotted to determine the path taken by the optimization procedure, as follows (see Figure 2):

```
R> plot(fit.drs, type = "trace")
```

The long stretch of nearly-constant parameter estimates and log-likelihood indicates a local maximum in the objective function. In general, a global optimum solution is not guaranteed with numerical techniques. The estimated baseline hazard with point-wise 95% bootstrapped confidence intervals is given by (see Figure 3):

```
R> set.seed(2015)
R> plot(fit.drs, type = "cumhaz", CI = 0.95)
```

where the seed is used to generate the weights in the bootstrap procedure of the cumulative baseline hazard plot function. Individual failures are shown by the rug plot directly above the time axis. Note that any other CI interval can be specified by the `CI` parameter of the `plot` method for 'fitfrail' objects. Subsequent calls to the `vcov` method for 'fitfrail' objects with the same arguments will use a cached value and avoid repeating the computationally-expensive bootstrap or sandwich variance estimation procedures.

For comparison, the following results were obtained with `coxph` in the **survival** package based on the PPL approach:

```
R> library("survival")
R> coxph(Surv(time, status) ~ treated + frailty.gamma(subject_id), drs)
```

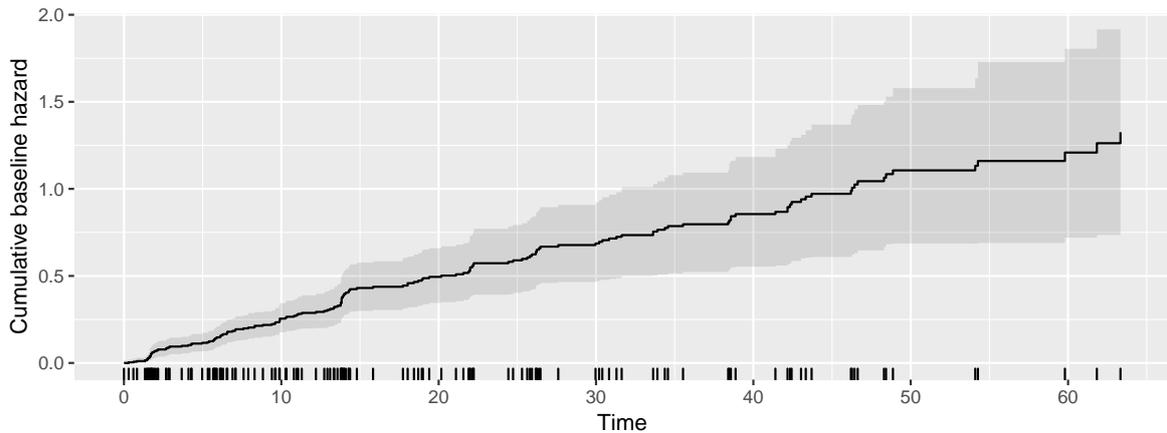

Figure 3: Estimated baseline hazard with point-wise 95% bootstrapped confidence intervals.

```
Call:
coxph(formula = Surv(time, status) ~ treated + frailty.gamma(subject_id),
    data = drs)

                              coef se(coef)     se2   Chisq   DF       p
treated                     -0.910    0.174   0.171  27.295  1.0 1.7e-07
frailty.gamma(subject_id)                            114.448 84.6   0.017

Iterations: 6 outer, 30 Newton-Raphson
     Variance of random effect= 0.854   I-likelihood = -850.9
Degrees of freedom for terms=  1.0 84.6
Likelihood ratio test=201   on 85.6 df, p=2.57e-11   n= 394
```

### 5.2. Hard drive failure

A dataset of hard drive monitoring statistics and failure was analyzed. Daily snapshots of a large backup storage provider over two years were made publicly available[4]. On each day, the Self-Monitoring, Analysis, and Reporting Technology (SMART) statistics of operational drives were recorded. When a hard drive was no longer operational, it was marked as a failure and removed from the subsequent daily snapshots. New hard drives were also continuously added to the population. In total, there are over 52,000 unique hard drives over approximately two years of follow-up and 2885 (5.5%) failures.

The data must be pre-processed in order to extract the SMART statistics and failure time of each unique hard drive. In some cases, a hard drive fails to report any SMART statistics up to several days before failing and the most recent SMART statistics before failing are recorded. The script for pre-processing is publicly available[5]. Although there are 40 SMART statistics altogether, many (older) drives only report a partial list. The current study is restricted to the covariates described in Table 2, which are present for all but one hard drive in the dataset.

---

[4] https://www.backblaze.com/hard-drive-test-data.html
[5] https://github.com/vmonaco/frailtySurv-jss

| Name | Description |
|------|-------------|
| `temp` | Continuous covariate, which gives the internal temperature in °C. |
| `rer` | Binary covariate, where 1 indicates a non-zero rate of errors that occur in hardware when reading from data from disk. |
| `rsc` | Binary covariate, where 1 indicates sectors that encountered read, write, or verification errors. |
| `psc` | Binary covariate, where 1 indicates there were sectors waiting to be remapped due to an unrecoverable error. |

Table 2: Hard drive failure covariates.

The hard drive lifetimes are thought to be clustered by model and manufacturer. There are 85 unique models ranging in capacity from 80 gigabytes to 6 terabytes. The cluster sizes loosely follow a power-law distribution, with anywhere from 1 to over 15,000 hard drives of a particular model.

For a fair comparison, the hard drives of a single manufacturer were selected. The subset of Western Digital hard drives consists of 40 different models with 178 failures out of 3530 hard drives. The hard drives are clustered by model, and cluster sizes range from 1 to 1190 with a mean of 88.25. A gamma shared frailty model was fitted to the data using the `"score"` fit method and default convergence criteria.

```
R> data("hdfail", package = "frailtySurv")
R> hdfail.sub <- subset(hdfail, grepl("WDC", model))
R> fit.hd <- fitfrail(
+    Surv(time, status) ~ temp + rer + rsc + psc + cluster(model),
+    hdfail.sub, frailty = "gamma", fitmethod = "score")
R> fit.hd

Call: fitfrail(formula = Surv(time, status) ~ temp + rer + rsc + psc +
    cluster(model), dat = hdfail.sub, frailty = "gamma", fitmethod = "score")

      Covariate       Coefficient
           temp           -0.0145
            rer            0.7861
            rsc            0.9038
            psc            2.4414

Frailty distribution    gamma(1.501), VAR of frailty variates = 1.501
Log-likelihood          -1305.134
Converged (method)      10 iterations, 15.78 secs (solved score equations)
```

Bootstrapped standard errors for the regression coefficients and frailty distribution parameter are given by

```
R> set.seed(2015)
R> COV <- vcov(fit.hd, boot = TRUE)
R> se <- sqrt(diag(COV)[c("temp", "rer", "rsc", "psc", "theta.1")])
R> se
```

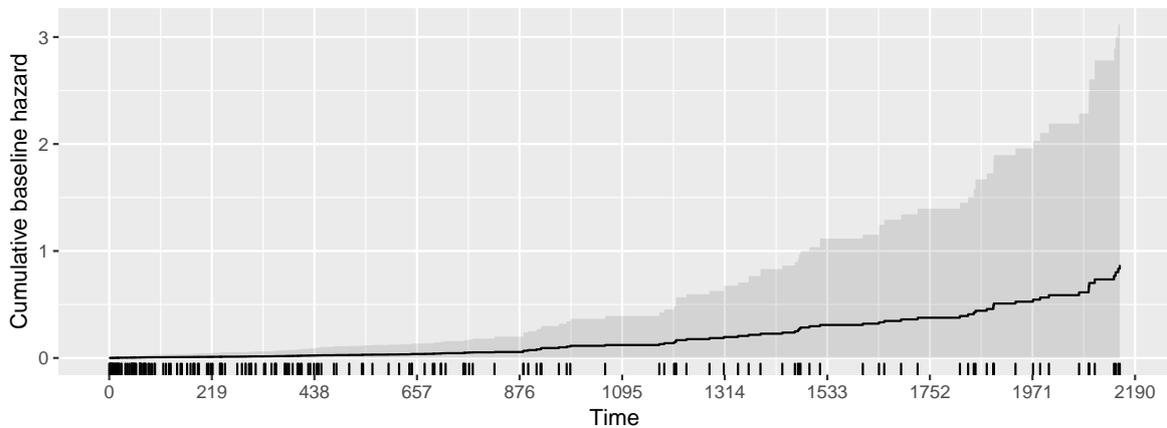

Figure 4: Estimated baseline hazard with 95% confidence interval.

```
      temp         rer         rsc         psc     theta.1
0.03095664  0.62533725  0.18956662  0.36142850  0.32433275
```

Significance of the regression coefficient estimates are given by their corresponding *p* values,

```
R> pvalues <- pnorm(abs(c(fit.hd$beta, fit.hd$theta)) / se,
+    lower.tail = FALSE) * 2
R> pvalues
```

```
         temp          rer          rsc          psc      theta.1
6.400162e-01 2.087038e-01 1.861996e-06 1.429667e-11 3.690627e-06
```

Only the estimated regression coefficients of the reallocated sector count (`rsc`) and pending sector count (`psc`) are statistically significant at the 0.05 level. Generally, SMART statistics are thought to be relatively weak predictors of hard drive failure (Pinheiro, Weber, and Barroso 2007). A hard drive is about twice as likely to fail with at least one previous bad sector (given by `rsc > 0`), while the hazard increases by a factor of 11 with the presence of bad sectors waiting to be remapped. The estimated baseline hazard with 95% CI is also plotted, up to 6 years, in Figure 4. This time span includes all but one hard drive that failed after 15 years (model: WDC WD800BB).

```
R> plot(fit.hd, type = "cumhaz", CI = 0.95, end = 365 * 6)
```

## 6. Discussion

**frailtySurv** provides a suite of functions for generating clustered survival data, fitting shared frailty models under a wide range of frailty distributions, and visualizing the output. The semi-parametric model has better asymptotic properties than most existing implementations, including consistent and asymptotically-normal estimators, which penalized partial likelihood estimation lacks. Moreover, this is the first R package that implements semi-parametric estimators with inverse Gaussian and PVF frailty models. The complete set of supported frailty distributions, including implementation details, are described in Appendix A. The flexibility and robustness of data generation and model fitting functions are demonstrated in Appendix B through a series of simulations.

The main limitation of **frailtySurv** is the computational complexity, which is approximately an order of magnitude greater than PPL. Despite this, critical sections of code have been optimized to provide reasonable performance for small and medium sized datasets. Specifically, **frailtySurv** caches computationally-expensive results, parallelizes independent computations, and makes extensive use of natively-compiled C++ functions through the **Rcpp** R package (Eddelbuettel and François 2011). As a remedy for relatively larger computational complexity, control parameters allow for fine-grained control over numerical integration and outer loop convergence, leading to a speed-accuracy tradeoff in parameter estimation.

The runtime performance and speed-accuracy tradeoff of core **frailtySurv** functions are examined empirically in Appendix C. These simulations confirm the $O(n)$ complexity of `genfrail` and $O(n^2)$ complexity of `fitfrail` using either log-likelihood maximization or normalized score equations. Frailty distributions without analytic Laplace transforms have the additional overhead of numerical integration inside the double-nested loop, although the growth in runtime is comparable to those without numerical integration. Covariance matrix estimation also has complexity $O(n^2)$, dominated by memory management and matrix operations. In order to obtain a tradeoff between speed and accuracy, the convergence criteria of the outer loop estimation procedure and convergence of numerical integration (for LN and IG frailty) can be specified through parameters to `fitfrail`. Accuracy of the regression coefficient estimates and frailty distribution parameter, as measured by the residuals, decreases as the absolute and relative reduction criteria in the outer loop are relaxed (Figure 18 in Appendix B). The simulations also indicate a clear reduction in runtime as numerical integration criteria are relaxed without a significant loss in accuracy (Figure 19 in Appendix B).

Choosing a proper frailty distribution is a challenging problem, although extensive simulation studies suggest that misspecification of the frailty distribution does not affect the bias and efficiency of the regression coefficient estimators substantially, despite the observation that a different frailty distribution could lead to appreciably different association structures (Glidden and Vittinghoff 2004; Gorfine, De-Picciotto, and Hsu 2012). There are several existing works on tests and graphical procedures for checking the dependence structures of clusters of size two (Glidden 1999; Shih and Louis 1995; Cui and Sun 2004; Glidden 2007). However, implementation of these procedures requires substantial extension to the current package, which will be considered in a separate work.

# Acknowledgments

The authors would like to thank Google, which partially funded development of **frailtySurv** through the *2015 Google Summer of Code*, and NIH grants (R01CA195789 and P01CA53996).

# A. Frailty distributions

All the frailty distributions used in **frailtySurv** have support $\omega \in (0, \infty)$. Identifiability problems are avoided by constraining the parameters when necessary. The gamma and PVF have a closed-form analytic expression for the Laplace transform, while the log-normal and inverse Gaussian Laplace transforms must be evaluated numerically. Analytic derivatives of the gamma and PVF Laplace transform were determined using the **Ryacas** R package (Goedman, Grothendieck, Højsgaard, Pinkus, and Mazur 2016). The resulting symbolic expressions were verified by comparison to numerical results. All the frailty distribution functions have both R and C++ implementations, while the C++ functions are used in parameter estimation. The **Rcpp** R package provides an interface to compiled native code (Eddelbuettel and François 2011). Numerical integration is performed by h-adaptive cubature (multi-dimensional integration over hypercubes), provided by the **cubature** C library (Johnson 2013), which implements algorithms described in Genz and Malik (1980) and Berntsen, Espelid, and Genz (1991).

For the gamma, log-normal, and inverse Gaussian, there is a positive relationship between the distribution parameter $\theta$ and the strength of dependence between cluster members. As $\theta$ increases, intra-cluster failure-times dependency increases. The opposite is true for the PVF, and as $\theta$ increases, the dependence between failure-times of the cluster's members decreases.

For frailty distributions with closed-form Laplace transforms, frailty variates are generated using a modified Newton-Raphson algorithm for numerical transform inversion (Ridout 2009). Note that while **frailtySurv** can generate survival data from a positive stable (PS) frailty distribution with Laplace transform $\mathcal{L}(s) = \exp(-\alpha s^\alpha/\alpha)$ where $0 < \alpha < 1$, it cannot estimate parameters for this model since the PS has infinite mean. Frailty values from a log-normal distribution are generated in the usual way, and inverse Gaussian variates are generated using a transformation method in the **statmod** package (Smyth, Hu, Dunn, Phipson, and Chen 2017).

## A.1. Gamma

Gamma distribution, denoted by $\text{Gamma}(\theta^{-1}) \equiv \text{Gamma}(\theta^{-1}, \theta^{-1})$, is of mean 1 and variance $\theta$. The **frailtySurv** package uses a one-parameter gamma distribution with shape and scale both $\theta^{-1}$, so the density function becomes

$$f(\omega; \theta) = \frac{\omega^{\frac{1}{\theta}-1} \exp\left(\frac{-\omega}{\theta}\right)}{\theta^{\frac{1}{\theta}} \Gamma(\frac{1}{\theta})} \ . \tag{25}$$

The special case with $\theta = 0$ is the degenerate distribution in which $\omega \equiv 1$, i.e., there is no unobserved frailty effect. Integrals in the log-likelihood function of Equation 20 can be solved using the Laplace transform derivatives, given by

$$\mathcal{L}^{(m)}(s) = (-1)^m \theta^{-\frac{1}{\theta}} \left(\theta^{-1} + s\right)^{-\left(\frac{1}{\theta}+m\right)} \Gamma\left(\theta^{-1} + m\right) / \Gamma\left(\theta^{-1}\right), \ \ m = 0, 1, 2, \dots, \tag{26}$$

where $\mathcal{L}^{(0)} = \mathcal{L}$. The first and second derivatives of the Laplace transform with respect to $\theta$ are also required for estimation. Due to their length, these expressions are omitted. See the `deriv_lt_dgamma_r` and `deriv_deriv_lt_dgamma_r` internal functions for the explicit expressions.

### A.2. Power variance function

The power variance function distribution is denoted by $\mathrm{PVF}(\theta, \delta, \theta)$ and with density

$$f(\omega; \theta, \delta, \mu) = \exp\left(-\mu\omega + \frac{\delta^\theta}{\theta}\right) \frac{1}{\pi} \sum_{k=1}^{\infty} \frac{\Gamma(k\theta + 1)}{k!} \left(-\frac{1}{\omega}\right)^{\theta k + 1} \sin(\theta k \pi), \qquad (27)$$

where $0 < \theta \leq 1, \mu \geq 0, \delta > 0$. To avoid identifiability problems, we let $\delta = \mu = 1$ as in Hanagal (2009), and get a one-parameter PVF density

$$f(\omega; \theta) = \exp\left(-\omega + \theta^{-1}\right) \frac{1}{\pi} \sum_{k=1}^{\infty} \frac{\Gamma(k\theta + 1)}{k!} \left(-\frac{1}{\omega}\right)^{\theta k + 1} \sin(\theta k \pi) . \qquad (28)$$

When $\theta = 1$, the degenerate distribution with $\omega \equiv 1$ is obtained. PVF has expectation 1 and variance $1 - \theta$. The Laplace transform is given by

$$\mathcal{L}(s) = \exp\left[-\left\{(1+s)^\theta - 1\right\}/\theta\right] . \qquad (29)$$

The Laplace transform derivatives are given by

$$\mathcal{L}^{(m)}(s) = (-1)^m L(s) \sum_{j=1}^{m} c_{m,j}(\theta) (1+s)^{j\theta - m}, \quad m = 1, 2, \ldots \qquad (30)$$

with coefficients

$$\begin{aligned}
c_{m,m}(\theta) &= 0 \\
c_{m,1}(\theta) &= \frac{\Gamma(m - \theta)}{\Gamma(1 - \theta)} \\
c_{m,j}(\theta) &= c_{m-1,j-1}(\theta) + c_{m-1,j}(\theta) \{(m-1) - j\theta\} .
\end{aligned}$$

The partial derivatives of the Laplace transform with respect to $\theta$ are given by

$$\begin{aligned}
\frac{\partial}{\partial \theta} \mathcal{L}^{(m)}(s) &= \frac{\partial}{\partial \theta} \left[(-1)^m \mathcal{L}(s) \sum_{j=1}^{m} c_{m,j}(\theta)(1+s)^{j\theta - m}\right] \\
&= (-1)^m \left\{\frac{\partial}{\partial \theta} \mathcal{L}(s)\right\} \sum_{j=1}^{m} c_{m,j}(\theta)(1+s)^{j\theta - m} \\
&\quad + (-1)^m \mathcal{L}(s) \sum_{j=1}^{m} \left\{\frac{\partial}{\partial \theta} c_{m,j}(\theta)(1+s)^{j\theta - m}\right. \\
&\qquad\qquad \left. + c_{m,j}(\theta) j (1+s)^{j\theta - m} \ln(1+s)\right\}, \qquad (31)
\end{aligned}$$

where

$$\frac{\partial}{\partial \theta} \mathcal{L}(s) = \exp\left\{\frac{1 - (s+1)^\theta}{\theta}\right\} \left\{-\frac{1 - (s+1)^\theta}{\theta^2} - \frac{(s+1)^\theta \log(s+1)}{\theta}\right\}$$

and the partial derivatives of the coefficients are

$$\begin{aligned}
\frac{\partial}{\partial \theta} c_{m,m}(\theta) &= 0 \\
\frac{\partial}{\partial \theta} c_{m,1}(\theta) &= \frac{\Gamma(m-\theta)\left\{\psi^{(0)}(1-\theta) - \psi^{(0)}(m-\theta)\right\}}{\Gamma(1-\theta)} \\
\frac{\partial}{\partial \theta} c_{m,j}(\theta) &= \frac{\partial}{\partial \theta} c_{m-1,j-1}(\theta) + \frac{\partial}{\partial \theta} c_{m-1,j}(\theta)\left\{(m-1) - j\theta\right\} - jc_{m-1,j}(\theta) .
\end{aligned}$$

### A.3. Log-normal

The log-normal distribution is denoted by LN($\theta$) and with density function

$$f(\omega;\theta) = \frac{1}{\omega\sqrt{\theta 2\pi}} \exp\left\{-\frac{(\ln \omega)^2}{2\theta}\right\}, \tag{32}$$

so the mean and variance are $\exp(\theta/2)$ and $\exp(2\theta) - \exp(\theta)$, respectively. The Laplace transform and its derivatives equal

$$\mathcal{L}^{(m)}(s) = \int_0^\infty (-\omega)^m e^{-s\omega} f(\omega;\theta)\, d\omega, \quad m = 0, 1, 2, \ldots . \tag{33}$$

Similar to the gamma distribution, the special case of $\theta = 0$ implies that $\omega \equiv 1$. The density's partial derivative with respect to $\theta$ is given by

$$\frac{\partial}{\partial \theta} f(\omega;\theta) = \frac{\ln^2(\omega) \exp\left(\frac{-\ln^2 \omega}{2\theta}\right)}{2\sqrt{2\pi}\theta^{5/2}\omega} - \frac{\exp\left(\frac{-\ln^2 \omega}{2\theta}\right)}{2\sqrt{2\pi}\theta^{3/2}\omega} . \tag{34}$$

### A.4. Inverse Gaussian

The inverse Gaussian distribution is denoted by IG($\theta$), with mean 1 and variance $\theta$. The density is given by

$$f(\omega;\theta) = \left(2\pi\theta\omega^3\right)^{-1/2} \exp\left\{\frac{-(\omega-1)^2}{2\theta\omega}\right\}, \tag{35}$$

where $\theta > 0$. The Laplace transform and its derivatives equal

$$\mathcal{L}^{(m)}(s) = \int_0^\infty (-\omega)^m e^{-s\omega} f(\omega;\theta)\, d\omega, \quad m = 1, 2, \ldots . \tag{36}$$

Similar to the gamma and log-normal, $\omega \equiv 1$ when $\theta = 0$. The partial derivative of the density function with respect to $\theta$ is given by

$$\frac{\partial}{\partial \theta} f(\omega;\theta) = \frac{(\omega-1)^2 \exp\left\{-\frac{(\omega-1)^2}{2\theta\omega}\right\}}{2\sqrt{2\pi}\theta^2\omega\sqrt{\theta\omega^3}} - \frac{\omega^3 \exp\left\{-\frac{(\omega-1)^2}{2\theta\omega}\right\}}{2\sqrt{2\pi}\left(\theta\omega^3\right)^{3/2}} .$$

# B. Simulation results

All simulations were run with 1000 repetitions, $n = 300$, fixed cluster size with $m = 2$ members within each cluster, covariates sampled from $\mathcal{U}(0, 1)$, regression coefficient vector $\beta = (\log 2, \log 3)^\top$, 30% censorship rate, and $\Lambda_0$ as in Equation 17 with $c = 0.01$ and $d = 4.6$, unless otherwise specified. The same seed is used for each configuration. Function calls are omitted for brevity and can be seen in the code repository[6].

## B.1. Benchmark simulation

As a benchmark simulation, we consider gamma frailty, with $\text{Gamma}(2)$. The results are summarized as follows:

```
Simulation: 1000 reps, 300 clusters (avg. size 2), gamma frailty
Serial runtime (s): 9812.27 (9.81 +/- 1.69 per rep)
         beta.1 beta.2 theta.1 Lambda.30 Lambda.60 Lambda.90
value    0.6931 1.0986  2.0000  0.003933   0.09539    0.6159
mean.hat 0.6821 1.0929  1.9752  0.003995   0.09716    0.6236
sd.hat   0.2472 0.2529  0.2659  0.001876   0.02248    0.1387
mean.se  0.3130 0.3156  0.3442        NA        NA        NA
cov.95CI 0.9890 0.9850  0.9780        NA        NA        NA
```

The cumulative baseline hazard true and estimated functions, with 95% point-wise confidence interval, is shown in Figure 5. Figure 6 indicates that by increasing the number of clusters, $n$, the bias and the variance of the estimators converge to zero, as expected.

## B.2. Large clusters

Increasing cluster size improves the estimated variances, especially of the frailty distribution parameter's estimator. The following simulation results are of $\text{Gamma}(2)$, $n = 100$ and fixed cluster size with $m = 6$, see also Figure 7.

```
Simulation: 1000 reps, 100 clusters (avg. size 6), gamma frailty
Serial runtime (s): 3665.11 (3.67 +/- 0.78 per rep)
         beta.1 beta.2 theta.1 Lambda.30 Lambda.60 Lambda.90
value    0.6931 1.0986  2.0000  0.003933   0.09539    0.6159
mean.hat 0.7077 1.0910  1.9979  0.003860   0.09554    0.6184
sd.hat   0.1958 0.2038  0.3100  0.001757   0.02191    0.1355
mean.se  0.2135 0.2182  0.3247        NA        NA        NA
cov.95CI 0.9620 0.9660  0.9530        NA        NA        NA
```

## B.3. Discrete observation times

Data generation allows for failure times to be rounded with respect to a specified base. The observed follow-up times were rounded to the nearest multiple of 10. The following simulation results indicate that even under the setting of ties, the empirical bias is reasonably small, and the empirical coverage rates of the confidence intervals are reasonably close to the nominal level. See the results below and Figure 8.

---
[6] https://github.com/vmonaco/frailtySurv-jss

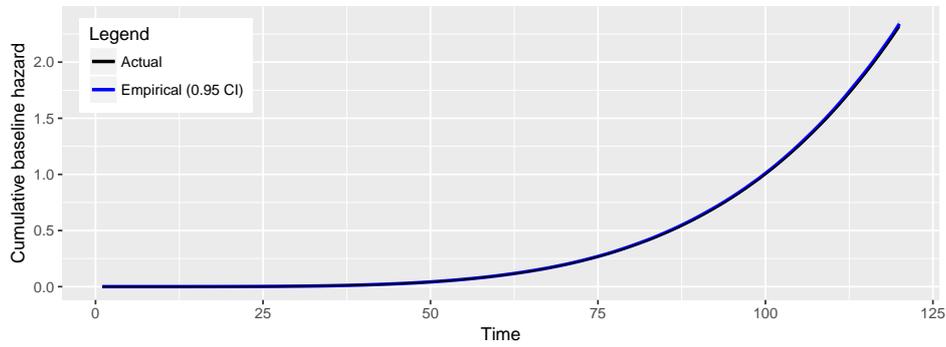

Figure 5: Cumulative baseline hazard true and estimated functions, with 95% point-wise confidence interval.

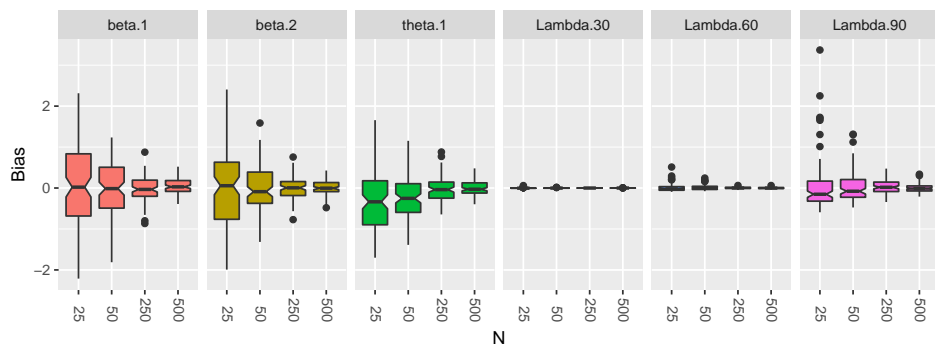

Figure 6: Distribution of the difference between estimated and true parameters in dependence of sample size.

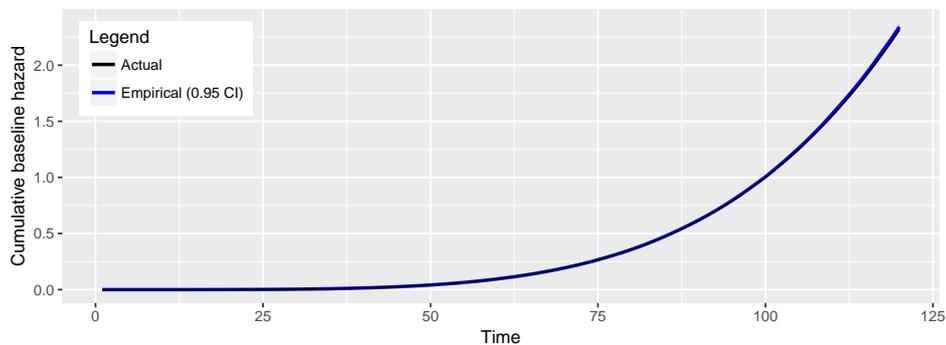

Figure 7: Cumulative baseline hazard true and estimated functions, with 95% point-wise confidence interval.

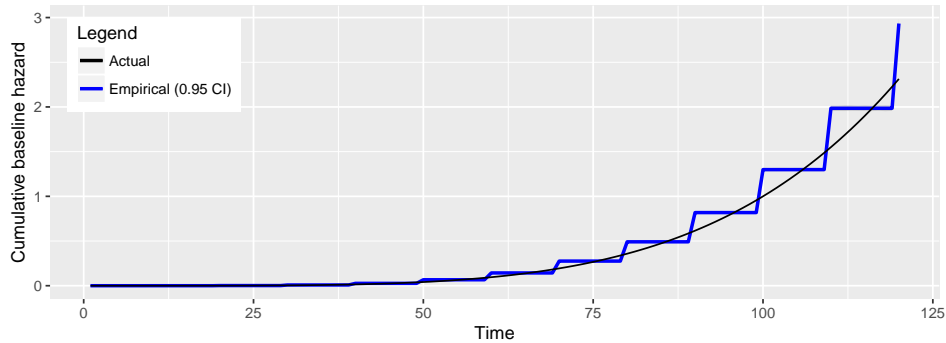

Figure 8: Cumulative baseline hazard true and estimated functions, with 95% point-wise confidence interval.

```
Simulation: 1000 reps, 300 clusters (avg. size 2), gamma frailty
Serial runtime (s): 10160.15 (10.16 +/- 1.73 per rep)
         beta.1 beta.2 theta.1 Lambda.30 Lambda.60 Lambda.90
value    0.6931 1.0986  2.0000  0.003933   0.09539    0.6159
mean.hat 0.6646 1.0736  1.9733  0.008242   0.14299    0.8182
sd.hat   0.2457 0.2487  0.2630  0.003010   0.03230    0.1837
mean.se  0.3127 0.3147  0.3454        NA        NA        NA
cov.95CI 0.9900 0.9850  0.9800        NA        NA        NA
```

### B.4. Oscillating baseline hazard

Consider the baseline hazard function

$$\lambda_0(t) = a^{\sin(b\pi t)} \left\{ d\left(ct\right)^d \right\} t^{-1} \quad t > 0 \tag{37}$$

where $a = 2$, $b = 0.1$, $c = 0.01$, and $d = 4.6$. Such an oscillatory component may be atypical in survival data, but demonstrates the flexibility of **frailtySurv** data generation and parameter estimation capabilities, as evident in the following simulation results (see also Figure 9).

```
Simulation: 1000 reps, 300 clusters (avg. size 2), gamma frailty
Serial runtime (s): 9560.76 (9.57 +/- 1.62 per rep)
         beta.1 beta.2 theta.1 Lambda.30 Lambda.60 Lambda.90
value    0.6931 1.0986  2.0000  0.005658   0.09050    0.7641
mean.hat 0.6810 1.0928  1.9747  0.005746   0.09226    0.7726
sd.hat   0.2462 0.2541  0.2656  0.002339   0.02150    0.1732
mean.se  0.3132 0.3157  0.3445        NA        NA        NA
cov.95CI 0.9880 0.9840  0.9830        NA        NA        NA
```

### B.5. Power variance function frailty

Power variance function frailty, with PVF (0.3) is considered, and the simulation results are summarized below and in Figure 10.

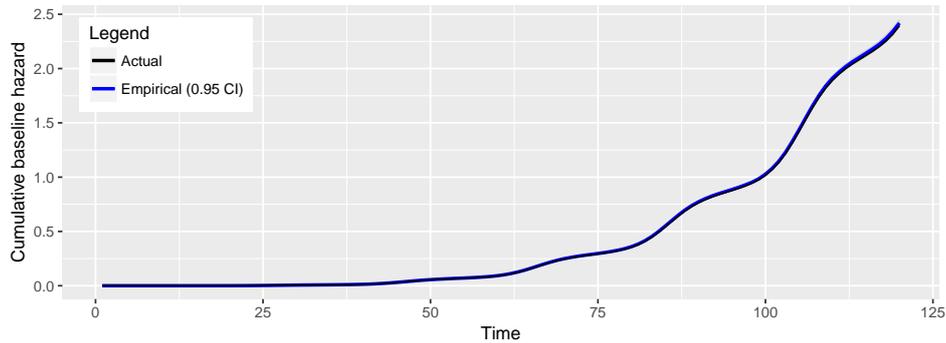

Figure 9: Cumulative baseline hazard true and estimated functions, with 95% point-wise confidence interval.

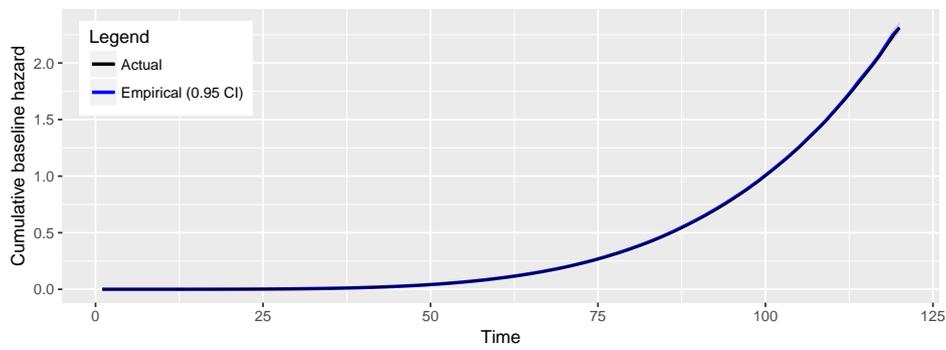

Figure 10: Cumulative baseline hazard true and estimated functions, with 95% point-wise confidence interval.

```
Simulation: 1000 reps, 300 clusters (avg. size 2), pvf frailty
Serial runtime (s): 9004.42 (9.00 +/- 2.01 per rep)
         beta.1 beta.2 theta.1 Lambda.30 Lambda.60 Lambda.90
value    0.6931 1.0986  0.3000  0.003933   0.09539    0.6159
mean.hat 0.6888 1.0899  0.3245  0.003990   0.09627    0.6223
sd.hat   0.2144 0.2127  0.1127  0.001756   0.01922    0.1124
mean.se  0.2643 0.2687  0.1266        NA        NA        NA
cov.95CI 0.9750 0.9880  0.9670        NA        NA        NA
```

### B.6. Poisson cluster sizes

Up until now, the cluster sizes have been held constant. Varying cluster sizes are typical in, e.g., geographical clustering and family studies. Consider the case in which the family size is randomly sampled from a zero-truncated Poisson with 2.313 mean family size. The following simulation results use PVF (0.3). The results are very good in terms of bias and the confidence intervals' coverage rates; see also Figure 11.

```
Simulation: 1000 reps, 300 clusters (avg. size 2.315), pvf frailty
```

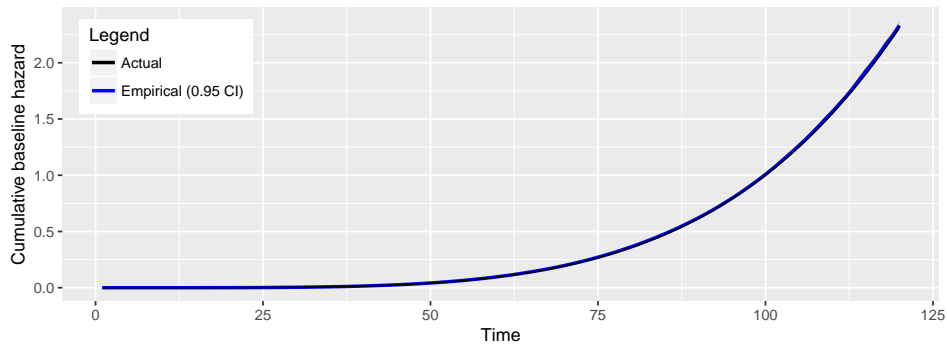

Figure 11: Cumulative baseline hazard true and estimated functions, with 95% point-wise confidence interval.

```
Serial runtime (s): 13383.31 (13.38 +/- 2.56 per rep)
         beta.1 beta.2 theta.1 Lambda.30 Lambda.60 Lambda.90
value    0.6931 1.0986 0.30000  0.003933   0.09539    0.6159
mean.hat 0.6830 1.0777 0.31944  0.004006   0.09767    0.6219
sd.hat   0.1837 0.2020 0.09878  0.001620   0.01838    0.1030
mean.se  0.2361 0.2411 0.10604        NA        NA        NA
cov.95CI 0.9870 0.9790 0.95700        NA        NA        NA
```

### B.7. Log-normal frailty

In this simulation, $LN(2)$ was used. The frailty variance equals 47.2. See results below and Figure 12.

```
Simulation: 1000 reps, 300 clusters (avg. size 2), lognormal frailty
Serial runtime (s): 68060.81 (68.06 +/- 15.07 per rep)
         beta.1 beta.2 theta.1 Lambda.30 Lambda.60 Lambda.90
value    0.6931 1.0986  2.0000  0.003933   0.09539    0.6159
mean.hat 0.6902 1.0794  1.9597  0.004173   0.09885    0.6282
sd.hat   0.2374 0.2416  0.3805  0.001634   0.02387    0.1280
mean.se  0.3402 0.3557  0.5066        NA        NA        NA
cov.95CI 0.9950 0.9900  0.9650        NA        NA        NA
```

### B.8. Inverse Gaussian frailty

Finally, we used $IG(2)$, where the frailty variance equals 2.

```
Simulation: 1000 reps, 300 clusters (avg. size 2), invgauss frailty
Serial runtime (s): 83183.12 (83.18 +/- 17.43 per rep)
         beta.1 beta.2 theta.1 Lambda.30 Lambda.60 Lambda.90
value    0.6931 1.0986  2.0000  0.003933   0.09539    0.6159
mean.hat 0.6898 1.0862  1.9489  0.004077   0.09648    0.6203
sd.hat   0.2280 0.2305  0.6226  0.001855   0.02108    0.1328
```

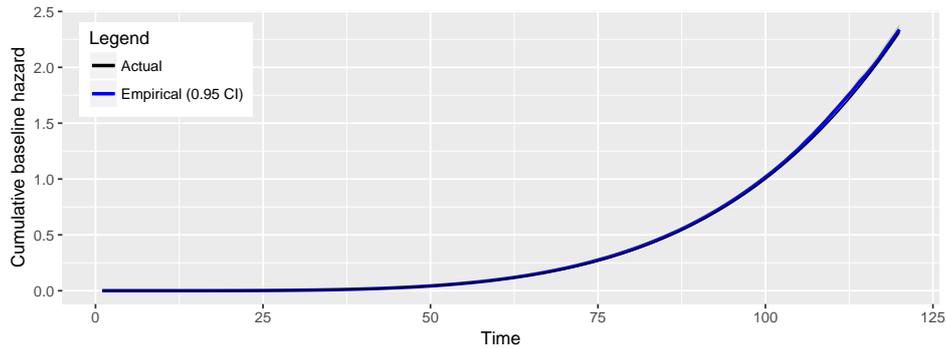

Figure 12: Cumulative baseline hazard true and estimated functions, with 95% point-wise confidence interval.

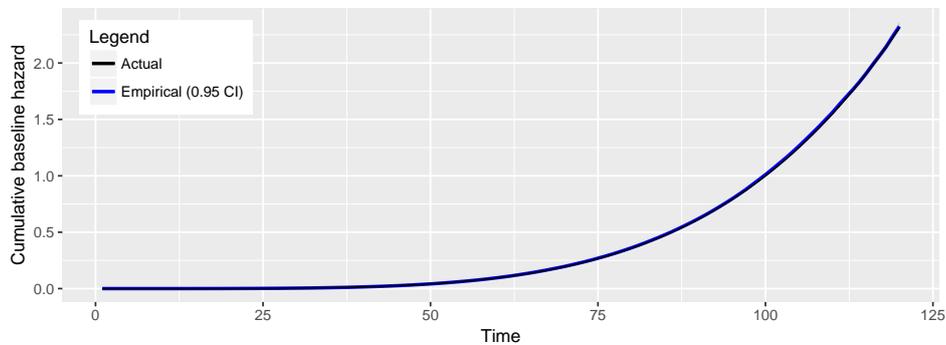

Figure 13: Cumulative baseline hazard true and estimated functions, with 95% point-wise confidence interval.

```
mean.se   0.2692 0.2685  0.8916          NA          NA          NA
cov.95CI  0.9840 0.9770  0.9520          NA          NA          NA
```

See also Figure 13.

## C. Performance analysis

Runtime was measured by the R function `system.time`, which measures the CPU time to evaluate an expression. All runs used 100 clusters of size 2, covariates sampled from $\mathcal{U}(0,1)$, regression coefficient vector $\beta = (\log 2, \log 3)^\top$, $\mathcal{N}(130, 15)$ censorship distribution, $\Lambda_0$ as in Equation 17 with $c = 0.01$ and $d = 4.6$, and 100 repetitions of each configuration, unless otherwise specified. The benchmark simulations were performed using a cluster of Red Hat 6.5 compute nodes, each with 2×2.6 GHz Intel Sandy Bridge (8 core) processors and 64 GB memory.

### C.1. Core functions

The runtimes of **frailtySurv** functions `genfrail` and `fitfrail`, and the `vcov` method for '`fitfrail`' objects were determined for increasing values of $n$, ranging from 50 to 200 in increments of 10. For each function, the runtime was determined for each of the four frailty distributions and each estimation procedure, where applicable. The bootstrap covariance runtime, i.e., `vcov` for '`fitfrail`' objects with `boot = TRUE`, was not analyzed since this consists primarily of repetitions of the parameter estimation function, `fitfrail`. The resulting runtimes are shown in Figures 14, 15, and 16, respectively.

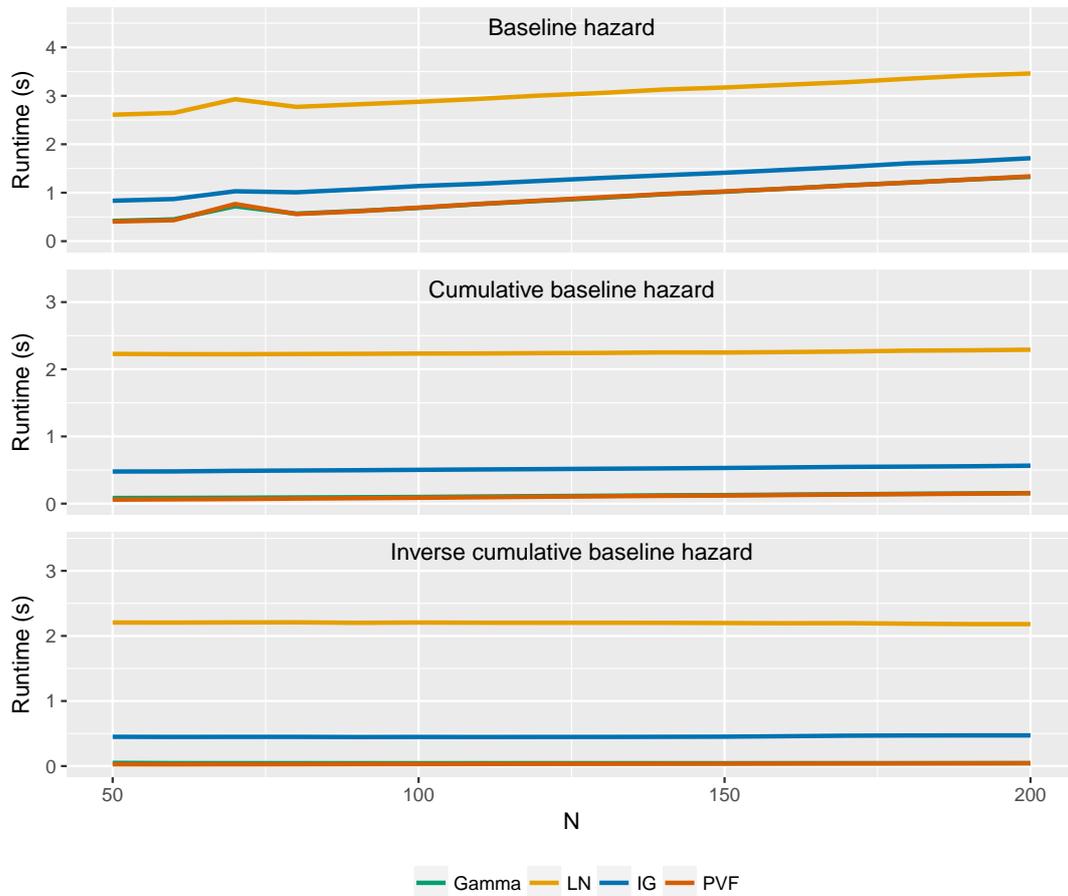

Figure 14: `genfrail` timings using each method of baseline hazard specification for increasing values of $n$. It is most efficient to specify the inverse cumulative baseline hazard to avoid solving for the root in Equation 5 and evaluating the integral in Equation 6.

Figure 14 shows the runtime of `genfrail`, which is linear in $n$, i.e., on the order of $O(n)$, although slope varies greatly depending on how the baseline hazard is specified. This is due to the amount of work that must be performed per observation. Specifying the cumulative baseline hazard or inverse cumulative baseline hazard to `genfrail` results in nearly-constant runtime. The linear increase in runtime is more apparent when the baseline hazard is specified since both root finding and numerical integration must be performed for each observation. The cumulative baseline hazard requires only root-finding to be performed, and the inverse cumulative baseline hazard has an analytic solution.

The runtimes of `fitfrail` using each estimation procedure and frailty distribution are shown in Figure 15. Both estimation procedures (log-likelihood reduction and normalized score equations) are on the order of $O(n^2)$ due to the doubly-nested loop. This complexity is more apparent for the log-normal and inverse Gaussian frailty distributions, which both have the additional overhead of numerical integration. Gamma and PVF frailty distributions have analytic Laplace transforms, thus numerical integration is not performed in the cumulative baseline hazard estimation inner loop.

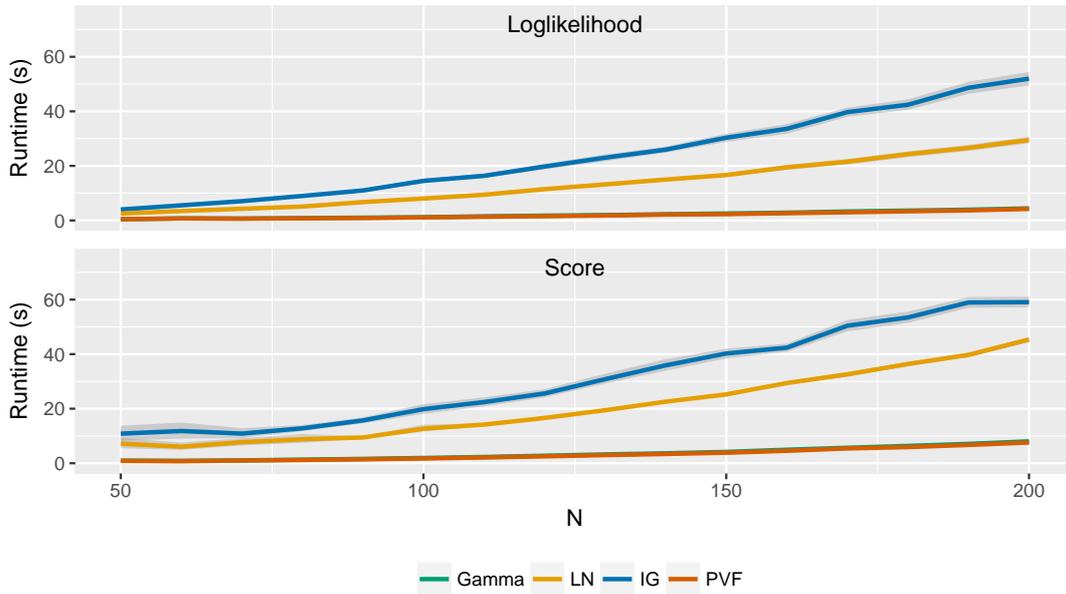

Figure 15: `fitfrail` timings using each estimation method for increasing values of $n$. The runtime for frailty distributions requiring numerical integration (inverse Gaussian and log-normal) grows quicker than those with analytic Laplace transforms (gamma and PVF).

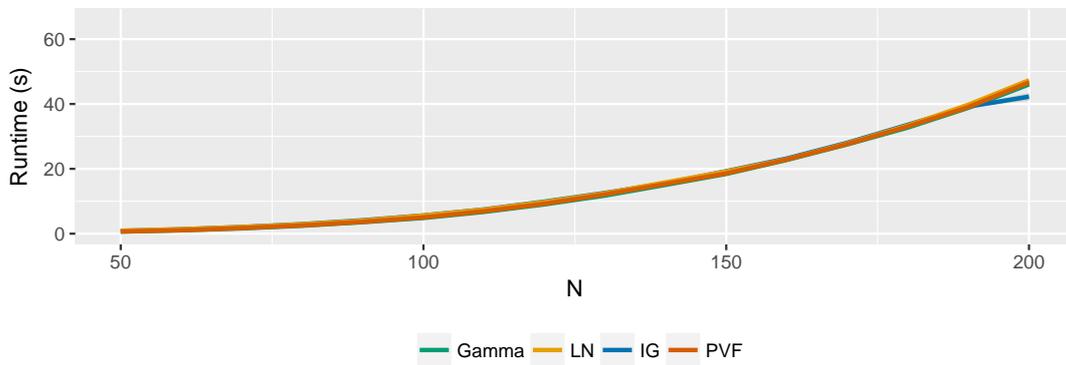

Figure 16: Timings for the `vcov` method for '`fitfrail`' objects for increasing values of $n$ using the analytic covariance estimator, i.e., with `boot = FALSE`.

Finally, the runtimes of the `vcov` method for '`fitfrail`' objects for each frailty distribution are shown in Figure 16. This function is also on the order of $O(n^2)$, and the sandwich variance estimation procedure is dominated by memory management and matrix operations to compute the Jacobian. As a result, the runtimes of frailty distributions that require numerical integration (LN and IG) are only marginally larger than those that do not (gamma and PVF).

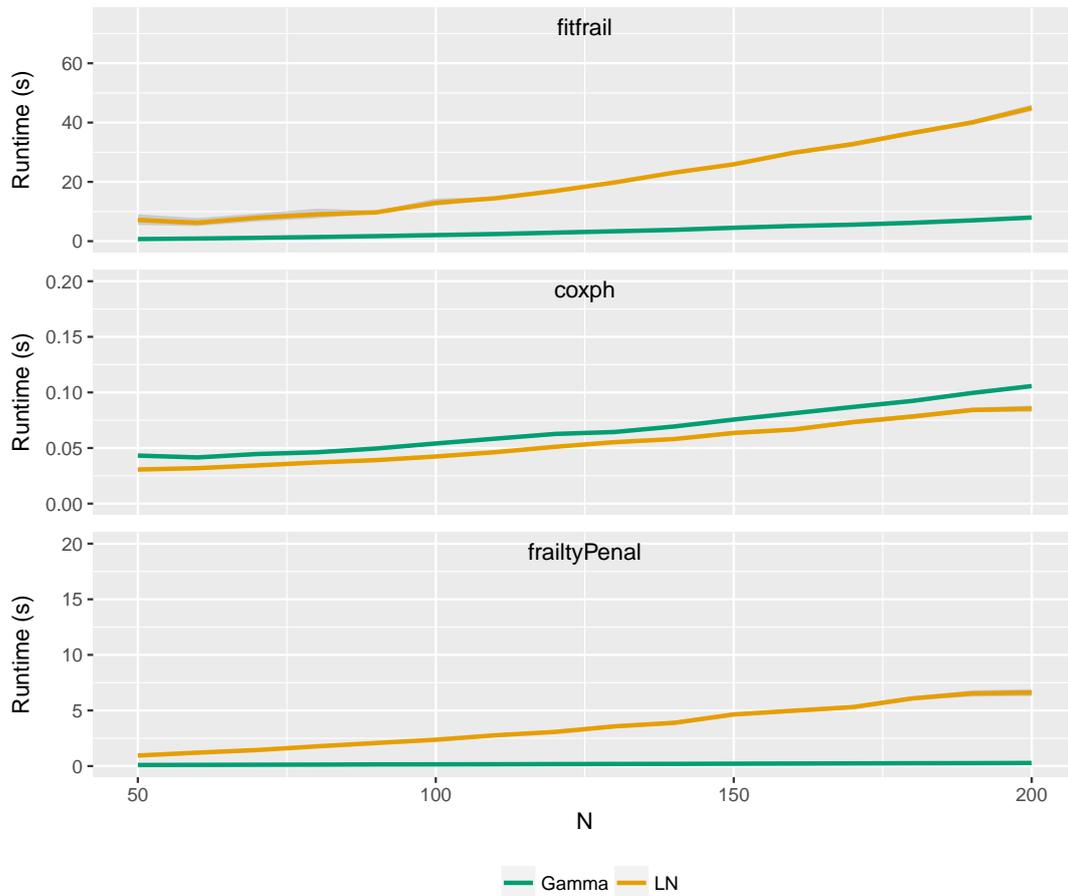

Figure 17: Comparison of frailty model estimation runtimes using **frailtySurv::fitfrail**, **survival::coxph**, and **frailtypack::frailtyPenal** for increasing values of $n$. `fitfrail` uses `fitmethod = "score"`, `coxph` uses default parameters, and `frailtyPenal` uses `n.knots = 10` and `kappa = 2`.

### C.2. Comparison to other packages

The runtime of `fitfrail` using `fitmethod = "score"` is compared to the functions `coxph` and `frailtyPenal` for increasing values of $n$. The runtimes for gamma and log-normal frailty distributions are determined, since this is the largest intersection of frailty distributions that all three functions support. The resulting runtimes are shown in Figure 17. Each estimation procedure exhibits quadratic complexity on a different scale. `coxph` remains roughly an order of magnitude quicker than `fitfrail` and `frailtyPenal` for log-normal frailty. `frailtyPenal` exhibits a large difference in performance between frailty distributions.

### C.3. Speed-accuracy tradeoff

A speed-accuracy tradeoff is achieved by varying the convergence control parameters of the outer loop estimation procedure and numerical integration in the inner loop. The `abstol` and `reltol` parameters control the outer loop convergence, and `int.abstol` and `int.reltol` control the convergence of the adaptive cubature numerical integration in the inner loop.

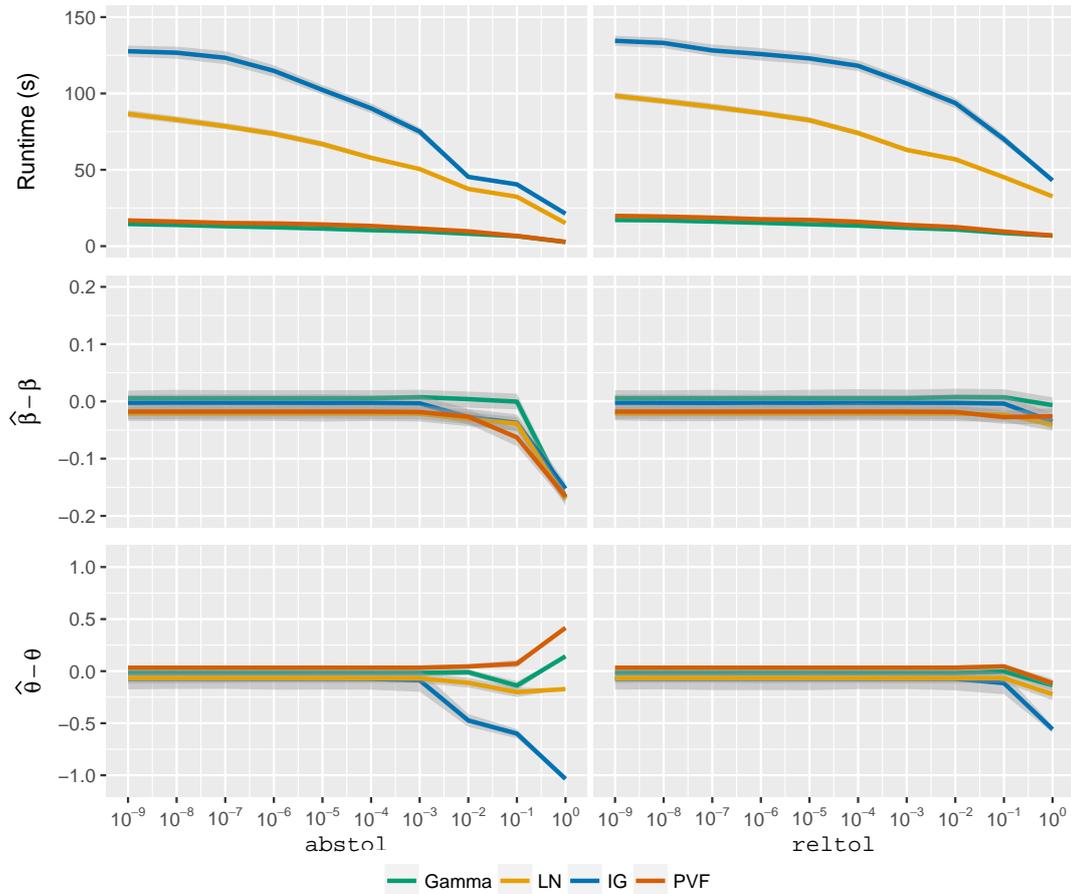

Figure 18: Speed and accuracy curves obtained by varying the outer loop convergence control parameters, `abstol` and `reltol`. As `abstol` is varied, `reltol` is set to 0 (i.e., it is ignored), and vice versa. $\widehat{\beta}-\beta$ and $\widehat{\theta}-\theta$ are the residuals of regression coefficient and frailty distribution parameter estimates, respectively. Shaded areas cover the 95% confidence intervals.

Speed is measured by the runtime of `fitfrail`, and accuracy is measured by the estimated parameter residuals.

In this set of simulations, the scalar regression coefficient $\beta = \log 2$, is used. Frailty distribution parameters are chosen such that $\kappa = 0.3$ ($\beta = 0.857$ for gamma, $\beta = 1.172$ for LN, $\beta = 2.035$ for IG, and $\beta = 0.083$ for PVF). With $\mathcal{N}(130, 15)$ right censorship distribution, this results in censoring rates 0.25 for gamma and PVF, 0.16 for LN, and 0.30 for IG. Both the runtime and residuals for $\beta$ and $\theta$ are reported using the `"score"` fit method.

Figure 18 shows the speed and accuracy curves for increasing values of `abstol` and `reltol` on a log scale, taking on values in $\{10^{-9}, \ldots, 10^0\}$. The runtime and residuals remain approximately constant up to $10^{-3}$ for both `abstol` and `reltol`. Beyond $10^{-3}$, the tradeoff between runtime and accuracy is apparent, especially for frailty distributions requiring numerical integration. As the convergence criterion is relaxed, runtime decreases and a bias is introduced to the parameter estimates. Runtimes for gamma and PVF frailty are nearly identical as these frailty distributions do not require numerical integration.

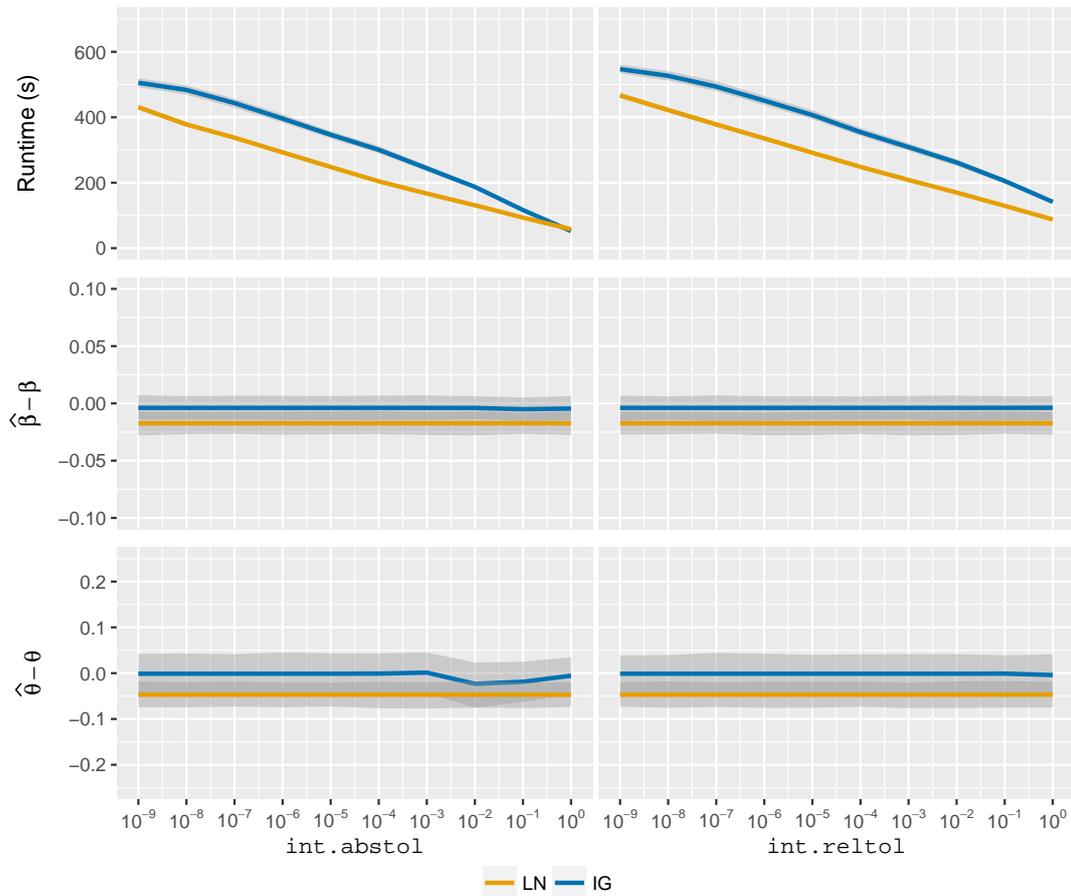

Figure 19: Speed and accuracy curves obtained by varying the inner loop numerical integration convergence control parameters, `int.abstol` and `int.reltol`. As `int.abstol` is varied, `int.reltol` is set to 0 (i.e., it is ignored), and vice versa. $\widehat{\beta} - \beta$ and $\widehat{\theta} - \theta$ are the residuals of regression coefficient and frailty distribution parameter estimates, respectively. Shaded areas cover the 95% confidence intervals.

Figure 19 shows the speed and accuracy curves obtained if the parameters `int.abstol` and `int.reltol` are varied over the same set of values for LN and IG frailty distributions. On this scale, the decrease in runtime is approximately linear, while residuals of the regression coefficient and frailty distribution parameters do not significantly change. We verified that the same behavior occurs using `fitmethod = "loglik"`. This suggests that the estimation procedure is robust to low-precision numerical integration with which significantly faster runtimes can be achieved. A strategy for parameter estimation on larger datasets might then be to first fit a model with a high value of `int.abstol` or `int.reltol` and iteratively decrease the numerical integration convergence until the parameter estimates do not change.

**Affiliation:**

John V. Monaco
Department of Computer Science
Naval Postgraduate School
Monterey, CA 93943, United States of America
E-mail: `vinnie.monaco@nps.edu`
URL: `http://www.vmonaco.com/`

Malka Gorfine
Department of Statistics and Operations Research
Tel Aviv University
Ramat Aviv
Tel Aviv, 6997801, Israel
E-mail: `gorfinem@post.tau.ac.il`
URL: `http://www.tau.ac.il/~gorfinem/`

Li Hsu
Public Health Sciences Division
Biostatistics and Biomathematics Program
Fred Hutchinson Cancer Research Center
1100 Fairview Ave. N., M2-B500
Seattle, WA 98109-1024, United States of America
E-mail: `lih@fredhutch.org`
URL: `https://www.fredhutch.org/en/labs/profiles/hsu-li.html`